\documentclass[%
 reprint,
showpacs,preprintnumbers,
 amsmath,amssymb,
 aps,
]{revtex4-1}

\usepackage{subcaption}
\usepackage{multirow}
\usepackage{xcolor}
\usepackage{float}
\usepackage{lipsum}
\usepackage{graphicx}
\usepackage{dcolumn}
\usepackage{bm}

\begin{document}

\preprint{APS/123-QED}

\title{Ultrahigh energy cosmic rays and neutrinos from light nuclei composition}

\author{Saikat Das$^1$}
\email{saikatdas@rri.res.in}
\author{Soebur Razzaque$^{2}$}
\email{srazzaque@uj.ac.za}
\author{Nayantara Gupta$^1$}
\email{nayan@rri.res.in}

\affiliation{$^1$ Astronomy \& Astrophysics Group, Raman Research Institute, Bengaluru 560080, India}

\affiliation{$^2$ Centre for Astro-Particle Physics (CAPP) and Department of Physics, University of Johannesburg, P.O. Box 524, Auckland Park 2006, South Africa}






\date{\today}

\begin{abstract}
The baryonic mass composition of ultrahigh energy ($\gtrsim 10^{18}$ eV) cosmic rays (UHECRs) at injection accompanied by their interactions on universal photon backgrounds during propagation directly governs the UHECR flux on the Earth. Secondary neutrinos and photons produced in these interactions serve as crucial astrophysical messengers of UHECR sources. A modeling of the latest data obtained by the Pierre Auger Observatory (PAO) suggests a mixed element composition of UHECRs with the sub-ankle spectrum being explained by a different class of sources than the super-ankle region ($> 10^{18.7}$ eV). In this work, we obtain two kinds of fit to the UHECR spectrum -- one with a single population of sources comprising of $^1$H and $^2$He, over an energy range commencing at $\approx 10^{18}$ eV -- another for a mixed composition of representative nuclei $^1$H, $^4$He, $^{14}$N and $^{28}$Si at injection, for which a fit is obtained from above $\approx 10^{18.7}$ eV. In both cases, we consider the source emissivity evolution to be a simple power-law in redshift. We test the credibility of H+He composition by varying the source properties over a wide range of values and compare the results to that obtained for H+He+N+Si composition, using the Monte Carlo simulation tool CRPropa 3. The secondary electrons and photons are propagated using the cosmic ray transport code DINT. We place limits on the source spectral index, source evolution index and cutoff rigidity of the source population in each case by fitting the UHECR spectrum. Cosmogenic neutrino fluxes can further constrain the abundance fraction and maximum source redshift in case of light nuclei injection model.

\end{abstract}

\maketitle


\section{\label{sec:intro}Introduction\protect}
The most powerful astrophysical accelerators in the Universe produce particles with energies at least up to few times $10^{20}$ eV. These are the highest energy particles observed in nature, called the ultrahigh energy cosmic ray (UHECR) particles with energies $E \gtrsim 10^{18}$ eV. Interpretation of the origin of UHECRs is a problem of foremost importance in astroparticle physics and a long-standing one \citep{linsley2, linsley1, sigl01, anchordoqui18}. Even after several decades of study, the nature and spatial distribution of sources, as well as the acceleration mechanism leading to the production of such high energy particles, remain elusive \cite{Kotera11}. The leading experiments to observe UHECRs are done at present by the Pierre Auger Observatory (PAO) in Argentina \citep{PAO1, PAO2} and the Telescope Array (TA) experiment in the United States \citep{TA1, TA2}. These experiments are expected to reach necessary sensitivity in upcoming years, that can unveil these mysteries.

UHECRs cannot be confined by the Galactic magnetic field at the highest energies, motivating for a search in extragalactic sources \citep{PAO3, hackstein}. The possible astrophysical sources include active galactic nuclei (AGNs) \citep{Berezinsky06, Dermer2009, wang17}, gamma-ray bursts (GRBs) \cite{Waxman:1995vg, zhang18}, low-luminosity GRBs \cite{Murase:2006mm}, hypernovae \cite{Wang:2007ya}, starburst galaxies \citep{anchordoqui_sb99, attallah18, anchordoqui_sb18}, gravitational accretion shocks \citep{kang96, berezinsky97}, neutron stars, etc. that can confine particles in their magnetic field up to a specific maximum energy $E_{\rm max}$ \cite{Hillas84}. 

UHECRs propagate through the Universe, interacting with the cosmic microwave background (CMB) and extragalactic background light (EBL). These interactions lead to interesting features in the UHECR spectrum at energies $E >10^{18}$ eV. The ankle at around $E \approx 5\times10^{18}$ eV, where hardening of the spectrum has been observed is assumed to be a feature resulting from the transition of Galactic to extragalactic cosmic rays \citep{aloisio12}. Another compelling possibility of interpreting the ankle is the $\mathrm{e^+}\mathrm{e^-}$ pair production dip caused by the interaction of cosmic ray protons with the CMB photons. The UHECR model with pure proton composition explaining the pair-production dip has been studied earlier in great detail \cite{Berezinsky06, Aloisio07, aloisio15}. The most prominent feature in the UHECR spectrum is the flux suppression at $E\gtrsim 5\times10^{19}$ eV, followed by a steep decline in the number of observed events. This may be a consequence of the interaction of UHECRs with CMB photons, called the GZK cutoff \cite{greisen, zatsepin} or, it can also be a manifestation of the maximum acceleration energy at the sources. UHECR interactions with the EBL photons become important below the GZK cutoff energy \cite{Dermer2009}. The uncertainties due to various EBL models is significant \cite{Batistaebl15, SimProp}. The Galactic and extragalactic magnetic fields deflect the UHECR particles during propagation. The deflection $\sim 10^\circ Z (40$ EeV $E^{-1})$ is higher for higher atomic number elements \citep{tinyakov05}. This makes a direct identification of UHECR sources impossible.

Charged and neutral pions are produced in UHECR interactions with the CMB and EBL, that decay to give high energy neutrinos and photons. Neutrinos are also produced from beta decay of neutrons. The flux of these cosmogenic neutrinos depends highly on the injection spectrum of UHECRs, the density and redshift evolution of the sources, the mass composition of UHECRs and also on the EBL models \citep{berezinsky69, engel01}. Unlike high energy photons and charged cosmic rays, high energy neutrinos can travel through cosmological distances unimpeded by interactions with other particles and undeflected by magnetic fields, providing a means to identify and study the extreme environments producing UHECRs. Thus, cosmogenic neutrinos are a definite probe to study UHECRs \citep{fang16b}. Current neutrino detectors \cite{Achterberg:2006md, Collaboration:2011nsa} have limited sensitivities to neutrinos at energies $>10^{16}$ eV but plans are underway to construct bigger and more sensitive experiments to detect such energetic neutrinos \cite{Aartsen:2014njl, Adrian-Martinez:2016fdl, ara, arianna, poemma1, grand1}. The diffuse gamma-ray background (DGRB) measured by Fermi-LAT \citep{ackermann15} can constrain the maximum cosmogenic photon flux produced from UHECR interactions and thus restrict UHECR source models. 

The mass-composition of UHECRs is not known to high precision. The atmospheric depth $X_{\rm max}$ where the number of particles in the cascade reaches its maximum is studied for this purpose \citep{gaisser13}. But reconstruction of $X_{\rm max}$ from shower simulations for an UHECR with given energy depends on the hadronic interaction models, which are uncertain at these extreme energies \citep{pierog15}. The average shower depth distribution $X_{\rm max}$, created by primary cosmic rays, indicates that UHECRs become heavier with increasing energy above $\approx 10^{18.2}$ eV \cite{Aab17}. Depending on the UHECR-air interaction model, this corresponds to a mass composition between p and He at $\approx 10^{18.2}$ eV and between He and N at $\approx 10^{19.5}$ eV. In some models, the $X_{\rm max}$ in the highest-energy bin ($10^{19.5}$--$10^{20}$ eV) is intermediate between N and Fe. The fluctuation in $X_{\rm max}$ that is $\sigma(X_{\rm max})$, which is another indicator of the mass composition, varies between H and He up to $10^{19.5}$ eV for all UHECR-air interaction models and is in between He and N in the $10^{19.5}$--$10^{20}$ eV energy bin \cite{Aab17}.    

Recent measurements at the PAO also shed light on the combined fit of the energy spectrum and mass composition data \cite{Aab17}, considering a simple astrophysical model of UHECR sources. The fit has been carried out for energies $> 5\times10^{18}$ eV, which is the region of all-particle spectrum above the ankle. Here, the ankle is interpreted as the transition between two (or more) different populations of sources. The astrophysical model assumes a homogeneous source distribution injecting five representative stable nuclei: H, He, N, Si, and Fe. The nuclei are accelerated through a rigidity-dependent mechanism. 
The results of the Auger fit puts forward a hard spectrum favoring low spectral indices ($E^{-\alpha}$ with $\alpha\lesssim 1$), making it difficult to conform with most particle acceleration models.

In a more recent study \cite{batista18}, the combined fit of the energy spectrum and mass composition made by the PAO is extended to some specific cases of source evolution corresponding to AGN, SFR, GRB, and power-law redshift dependence. The latter has a form $(1+z)^m$, where $m$ is a free parameter and results in slightly better fits. They obtained best fits for hard spectral indices ($\alpha\lesssim$1.0) and low maximal rigidities ($R_{\rm max}<10^{19}$ eV) for compositions at injection dominated by intermediate-mass nuclei (nitrogen and silicon groups). They show that negative source emissivity evolution is preferred, with the best fit for $m=-1.6$, ensuing hardest spectral indices provide the lowest possible cosmogenic fluxes for the $(1+z)^m$ source redshift evolution.

In this paper, we model the latest UHECR spectrum using two different astrophysical conditions. We analyze the ``CTD'' propagation model, mentioned in \citep{Aab17} using CRPropa 3 and Dom\'{i}nguez et al. EBL model \citep{dominguez11}. First, we consider a single population of sources, injecting only H and He. We fit the spectrum starting from $E \approx 10^{18}$ eV up to the highest energy data point observed by the PAO. In this case, the ankle is explained by $\mathrm{e^+}\mathrm{e^-}$ pair production of UHECRs on background photons. After fitting UHECR data, we discuss whether cosmogenic neutrinos can constrain the mass composition near the ankle, as well as other model parameters such as the redshift distribution of sources and the maximum UHECR energy. We present a technique to probe the mass composition at injection by future measurement of individual neutrino flavor fluxes. We vary the abundance fraction of injected elements to obtain a fit from $E\approx 10^{18}$ eV. A study of the correlation between fit parameters is done to reject unrealistic cases. Next, we consider another scenario, where the sources inject H, He, N, and Si to fit the UHECR spectrum for $E>10^{18.7}$ eV. A separate population of sources would be required to fit the spectrum for $E<10^{18.7}$ eV and above the knee in this scenario. We compare the results obtained for these two types of composition at injection to distinguish between favorable scenarios.

We discuss UHECR propagation, interactions and fluxes in general in Sec.\ \ref{sec:prop} and in details in Sec.\ \ref{sec:setup} for CRPropa 3.  Our results are presented in Sec.\ \ref{sec:result} and discussed in Sec.\ \ref{sec:discuss}. We draw our conclusions in Sec.\ \ref{sec:conclu}.

\section{\label{sec:prop}UHECR propagation and\\
cosmogenic fluxes\protect}

UHECRs propagate through the intergalactic space and interact with the cosmic background photons primarily via the $\Delta$-resonance channel. They lose their energy through secondary particle productions as
\begin{equation}
p+\gamma_{bg} \rightarrow \Delta^+ \rightarrow
\begin{cases}
p + \pi^0 \\
n + \pi^+
\end{cases}
\end{equation}
The neutral pion decays to give photons ($\pi^0\rightarrow\gamma\gamma$), and the charged pion decays to produce neutrinos ($\pi^+\rightarrow\mu^+ + \nu_\mu \rightarrow \mathrm{e^+} + \nu_e + \overline{\nu}_\mu$ + $\nu_\mu$). Additionally, there can be double pion production and multipion production processes with much lower cross sections \cite{anita00, murase06}. These photonuclear processes lead to the formation of the GZK feature in the cosmic ray spectrum, thereby causing a sharp decay at $E>5\times10^{19}$ eV \cite{greisen, zatsepin}, near to the threshold for the photopion production with the CMB photons ($E_{\rm th} \approx 6.8\times 10^{19}$ eV). Cosmic rays interact dominantly with low energy CMB photons of energy $\epsilon \sim 10^{-3}$ eV, during their propagation. EBL photons have energy higher than the CMB photons. This allows protons of energy lower than the threshold of photopion production with the CMB to interact with the EBL photons and generate neutrinos. Although the number of EBL photons is much smaller than the CMB, they have a significant effect on the neutrino flux. The CMB photon density increases with redshift as $(1+z)^3$. The spectral shape and cosmological evolution of the infrared, ultraviolet and optical backgrounds comprising the EBL are not as well known as the CMB. With the redshift evolution of the photon background, the interaction length of cosmic rays also evolves with redshift. 

Beta decay contributes to cosmogenic neutrino flux through the decay of neutrons resulting from the charged pion production,
\begin{equation}
n\rightarrow p+\mathrm{e^-} + \overline{\nu}_e
\end{equation}
Heavier nuclei with a higher atomic number ($Z>1$) also undergo beta decay and give rise to photopion production. Photodisintegration of nuclei due to irradiation by photons of energy between 8 to 30 MeV is the dominant energy loss mechanism for UHECR nuclei,
\begin{equation}
^A_ZX + \gamma \rightarrow ^{A-n}_{Z-n'}X + nN
\end{equation} 
In this process, a nucleus interacts inelastically with a cosmic background photon which leads to partial fragmentation of the nucleus producing $n (n')$ stripped nucleons (protons). The deexcitation of an excited nucleus can give high energy photons. UHECRs can also undergo Bethe-Heitler pair production to generate $\mathrm{e^+}\mathrm{e^-}$ pairs. Electron pair production has the largest cross section among the photohadronic interactions, the threshold energy being 2 orders of magnitude smaller than that of pion production. The electrons and positrons produced in various processes can induce electromagnetic cascades down to GeV energies and thus contribute to the cosmogenic photon flux. 

The energy loss rate of protons with energy $E$ due to cosmic expansion is expressed as,
\begin{equation}
\dfrac{dE}{dt}=-\dfrac{\dot{a}}{a}E=-H_0 \bigg[\Omega_m (1+z)^3 + \Omega_\Lambda\bigg]^{1/2}E
\end{equation}
where $a$ is the scale factor and $\Lambda$CDM cosmology is considered here with $H_0=67.3$ km s$^{-1}$ Mpc$^{-1}$, $\Omega_m=0.315$, $\Omega_\Lambda=1-\Omega_m$ \cite{olive}. Neutrinos, being weakly interacting, propagate unhindered through the cosmos and experience only adiabatic energy loss due to the cosmic expansion. The photons interact with cosmic background radiations and the universal radio background (URB) to produce electromagnetic cascades through various processes such as Breit-Wheeler pair production, double pair production, resulting in $\mathrm{e^+}\mathrm{e^-}$ pairs \citep{heitercrp}. The relativistic cascade electrons lose energy by triplet pair production, synchrotron radiation on deflection in magnetic fields and up-scattering background photons by inverse Compton scattering.

\section{\label{sec:setup}Setup for CRPropa simulations\protect}

CRPropa 3 is a public astrophysical simulation framework to propagate ultrarelativistic particles from their sources to the observer through the Galactic and extragalactic space. Primary and secondary cosmic messengers such as protons, pions, nuclei, charged leptons, neutrinos, and photons are produced as output \citep{CRPropa3, heitercrp}. We use CRPropa 3 to propagate primary and secondary UHECR protons and nuclei to get the particle yields obtained at the Earth. Secondary neutrinos produced by photopion production of UHECRs on background photons and beta decay of neutrons are also propagated. The secondary electromagnetic particles generated in ultrahigh energy nuclei propagation are stored and then propagated using the cosmic ray transport code DINT as an external program from within CRPropa \citep{lee98}. We use a uniform extragalactic magnetic field of strength $0.1$ nG for DINT propagation.

We include all possible energy loss processes for primary UHE protons and nuclei in the simulation, viz. photopion production, Bethe-Heitler pair production, photodisintegration (for $Z>1$), nuclear decay and adiabatic energy losses due to the expansion of the Universe. We assume an injection spectrum of primary particles at the UHECR source of the following form,
\begin{equation}
\dfrac{dN}{dE} = A_0 \sum_i K_i E^{-\alpha} \times f_{\text{cut}}(E, ZR_{\text{cut}})
\label{eq:injection}
\end{equation}
where $K_i$ is the abundance fraction of the $i-$th nuclei at injection, $E$ is the energy of the injected particle, $A_0$ is an arbitrary normalization flux, $\alpha$ is the spectral index, $Z$ is the charge of the primary cosmic ray and $R_{\text{cut}}=E_{\text{cut}}/Z$ is the cutoff rigidity. We use a broken exponential cutoff function in the injection spectrum given by,
\begin{align}
f_{\text{cut}}(E, ZR_{\text{cut}}) =
\begin{cases}
1 & \text(E<ZR_{\text{cut}})\\
\exp\bigg(1-\dfrac{E}{Z R_{\text{cut}}}\bigg) & \text(E>ZR_{\text{cut}})
\end{cases}
\end{align}
We assume particles are injected with energies between $E_{\rm min}=0.1$ EeV and $E_{\rm max}=1000$ EeV. We consider the evolution of source emissivity to be a simple power-law in redshift, given by $(1+z)^m$, where $m$ is a free parameter.

\begin{table}
\caption{\label{tab:param}UHECR parameters used for simulations}
\begin{ruledtabular}
\begin{tabular}{lll}
\textbf{Parameter} & \textbf{Description} & \textbf{Values}\\ 
\hline \\
$\alpha$ & Source spectral index & $2.2 \leqslant \alpha \leqslant 2.6
$ \\ 
$R_{\text{cut}}$ & Cutoff rigidity & $40 \leqslant R_{\text{cut}} \leqslant 100$ EV\\
$z_{\text{min}}$ & Minimum redshift &  $z_{\text{min}}=0.0007$ \\
$z_{\text{max}}$ & Cutoff redshift & $2\leqslant z_{\text{max}} \leqslant 4$\\
$m$ & Source evolution index & $0\leqslant m \leqslant3$\\
$K_i$ & Abundance fraction & $0.0$\% $\leqslant K_i < 100$\%\\
$A_0$ & Flux normalisation & $A_0 > 0$
\end{tabular}
\end{ruledtabular}
\end{table}

At ultrahigh energies, the cosmic rays interact with background radiation comprised of CMB and EBL. The CMB spectrum is well known to high precision and can be characterized by an isotropic blackbody spectrum with $T \approx 2.73$ K \cite{planck15}. The EBL models implemented in CRPropa 3 are Kneiske et al.~\cite{kneiske04}, Stecker et al.~\cite{stecker06}, Franceschini et al.~\cite{frances08}, Finke et al.~\cite{finke10}, Dom\'{i}nguez et al.~\cite{dominguez11}, Gilmore et al.~\cite{gilmore12} and also the upper and lower bounds determined by Stecker et al.~\cite{steckerup}. As we explore the plausible mass composition within the ``CTD'' model, the energy loss interactions of UHECRs and photon backgrounds (CMB and EBL) are considered with the TALYS 1.8 photodisintegration model \cite{koning05} and the Dom\'{i}nguez et al.~\cite{dominguez11} EBL model. Interactions with the magnetic field are relevant for charged particles, mainly electrons, and positrons produced in electromagnetic cascades and are taken into account for DINT propagation. UHECR protons and nuclei being much heavier than electrons have no significant energy loss in magnetic field interactions. Since we are interested in composition and energy spectrum study, we consider a null Galactic and extragalactic magnetic field for UHECR propagation. Hence, our simulations are effectively one-dimensional.

\section{\label{sec:result}Results\protect}

\begin{figure*}[hbt]
\centering
\begin{subfigure}{.5\textwidth}
  \centering
  \includegraphics[width=8.0 cm, height = 5.2 cm]{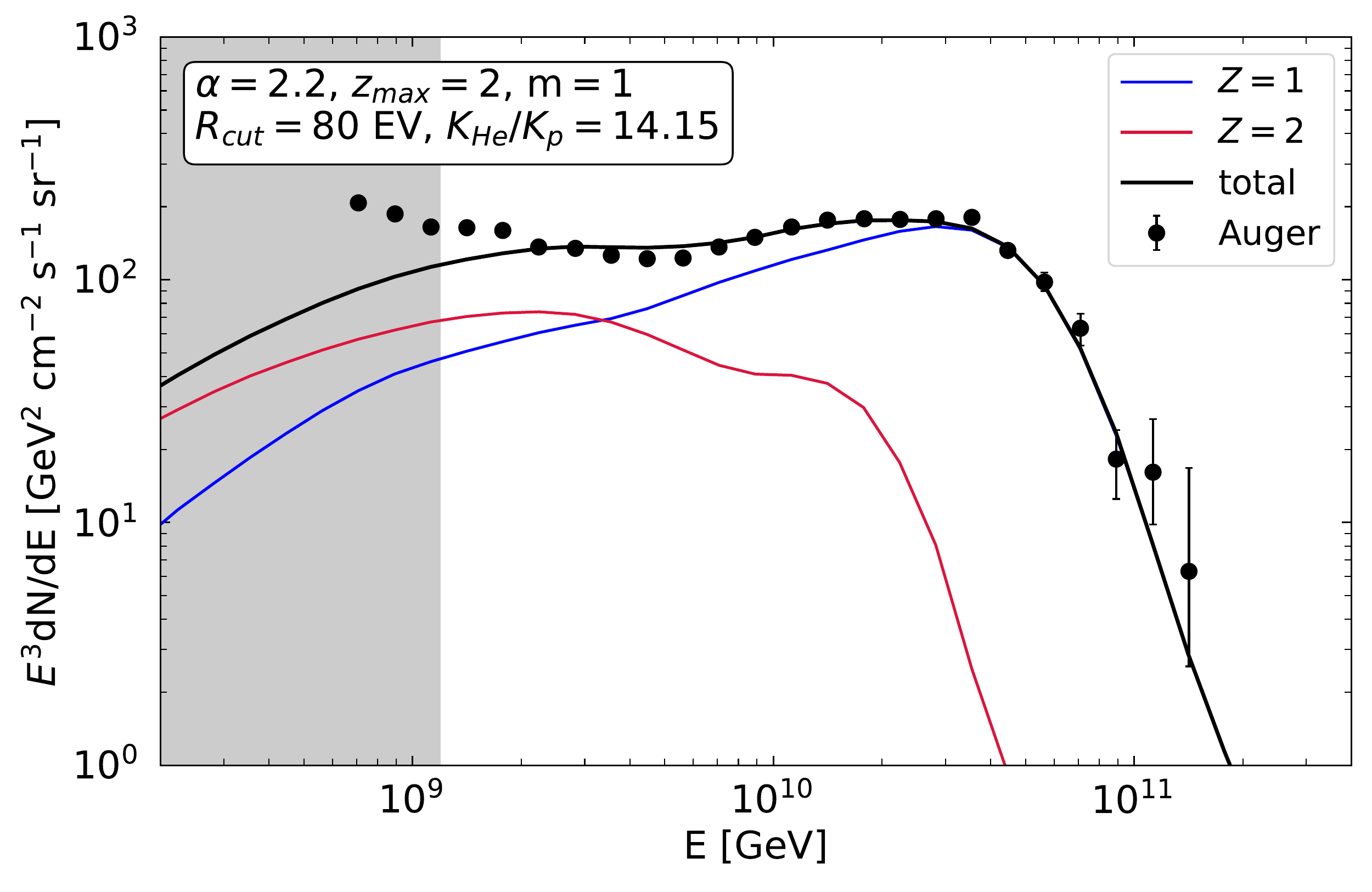}
\end{subfigure}%
\begin{subfigure}{.5\textwidth}
  \centering
  \includegraphics[width=8.0 cm, height = 5.2 cm]{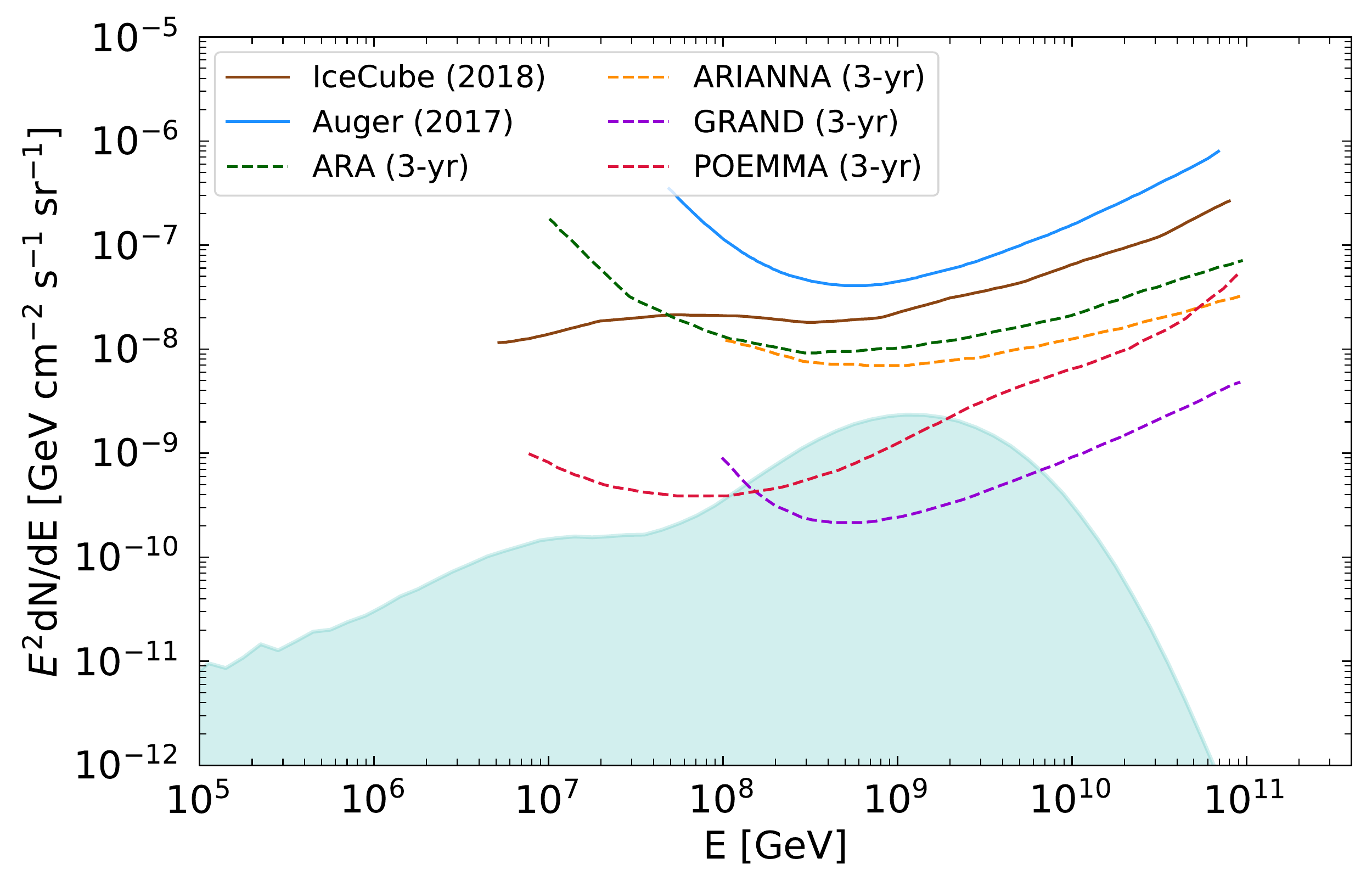}
\end{subfigure}
\begin{subfigure}{.5\textwidth}
  \centering
  \includegraphics[width=8.0 cm, height = 5.2 cm]{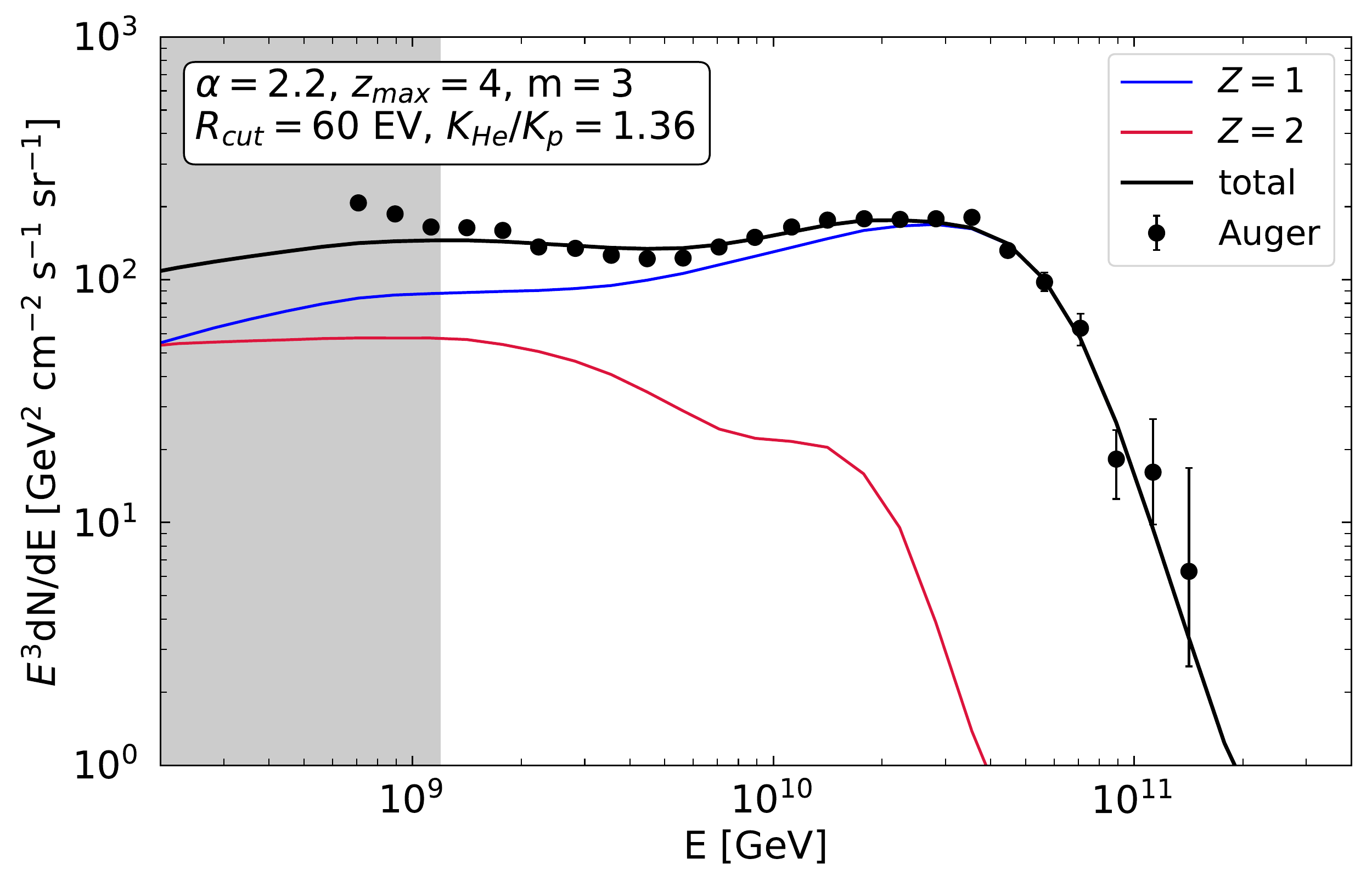}
\end{subfigure}%
\begin{subfigure}{.5\textwidth}
  \centering
  \includegraphics[width=8.0 cm, height = 5.2 cm]{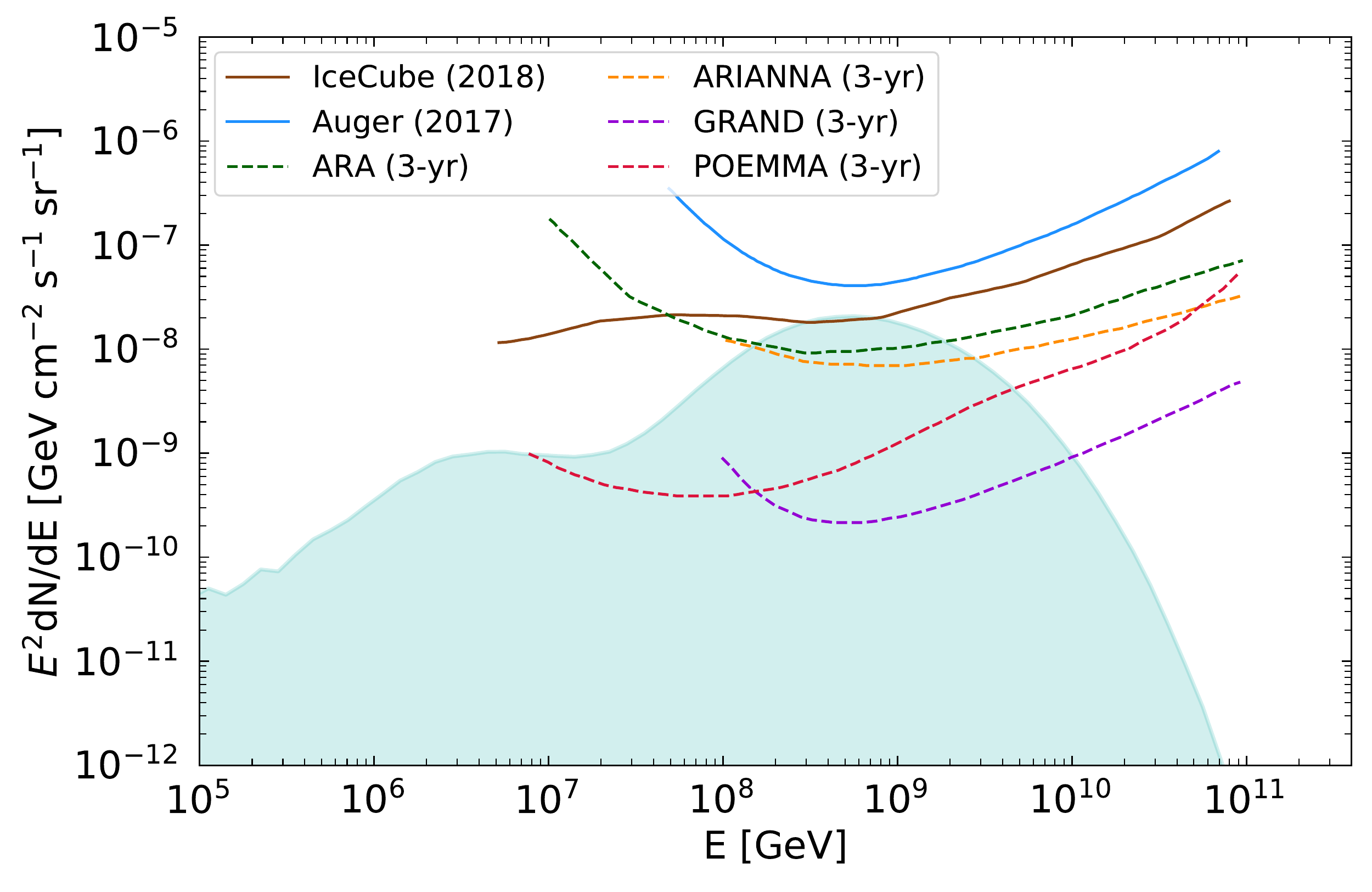}
\end{subfigure}
\caption{\small{UHECR spectra (left) and cosmogenic neutrino spectra (right) for $\alpha=2.2$. The top (case 2) and bottom (case 12) panels show the best-fit cases listed in Appendix A for which the difference in the cosmogenic neutrino flux is the maximum.}}
\label{fig:2.2}
\end{figure*}

\begin{figure*}[hbt]
\begin{subfigure}{.5\textwidth}
  \centering
  \includegraphics[width=8.0 cm, height = 5.2 cm]{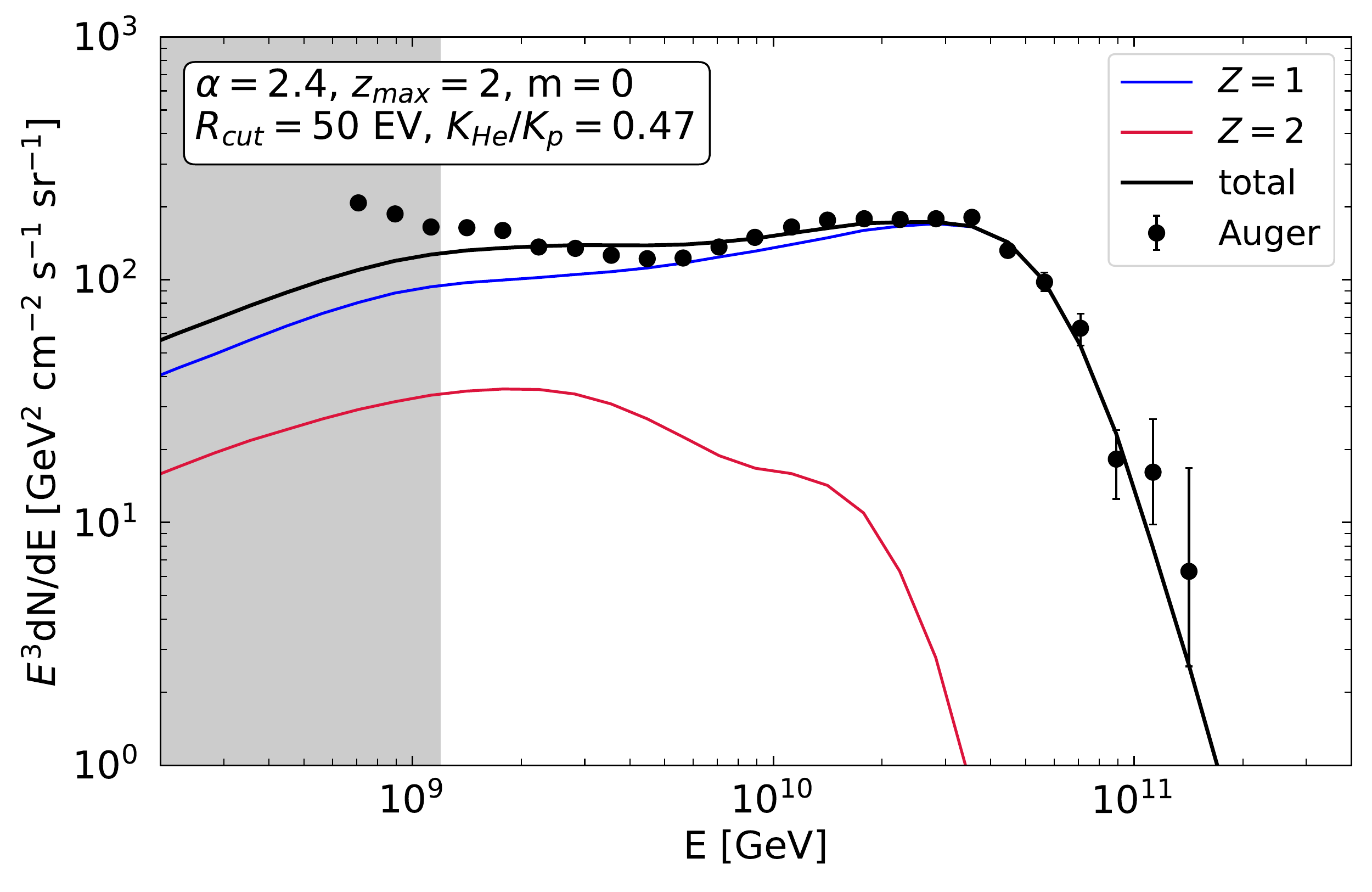}
\end{subfigure}%
\begin{subfigure}{.5\textwidth}
  \centering
  \includegraphics[width=8.0 cm, height = 5.2 cm]{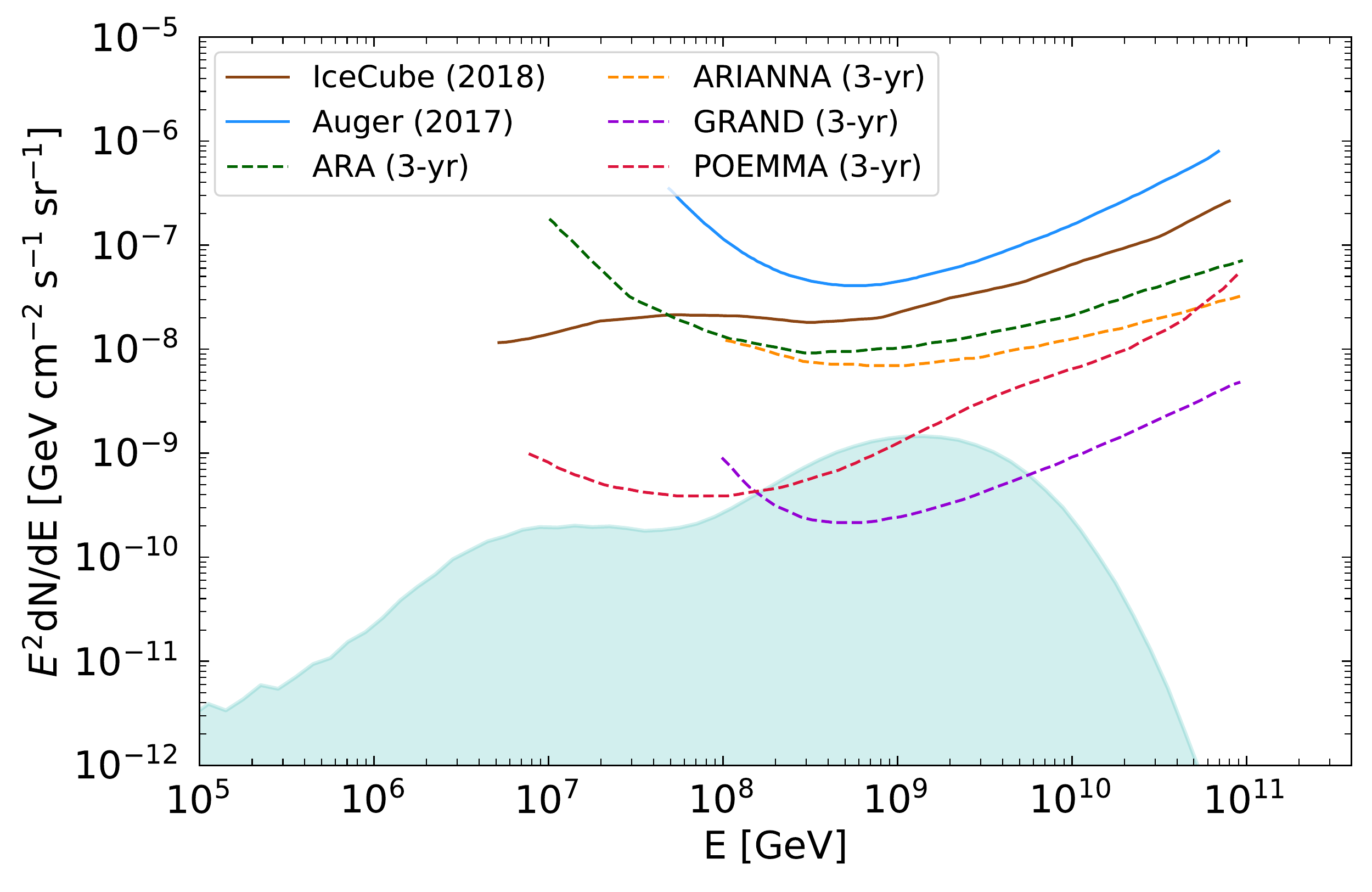}
\end{subfigure}
\begin{subfigure}{.5\textwidth}
  \centering
  \includegraphics[width=8.0 cm, height = 5.2 cm]{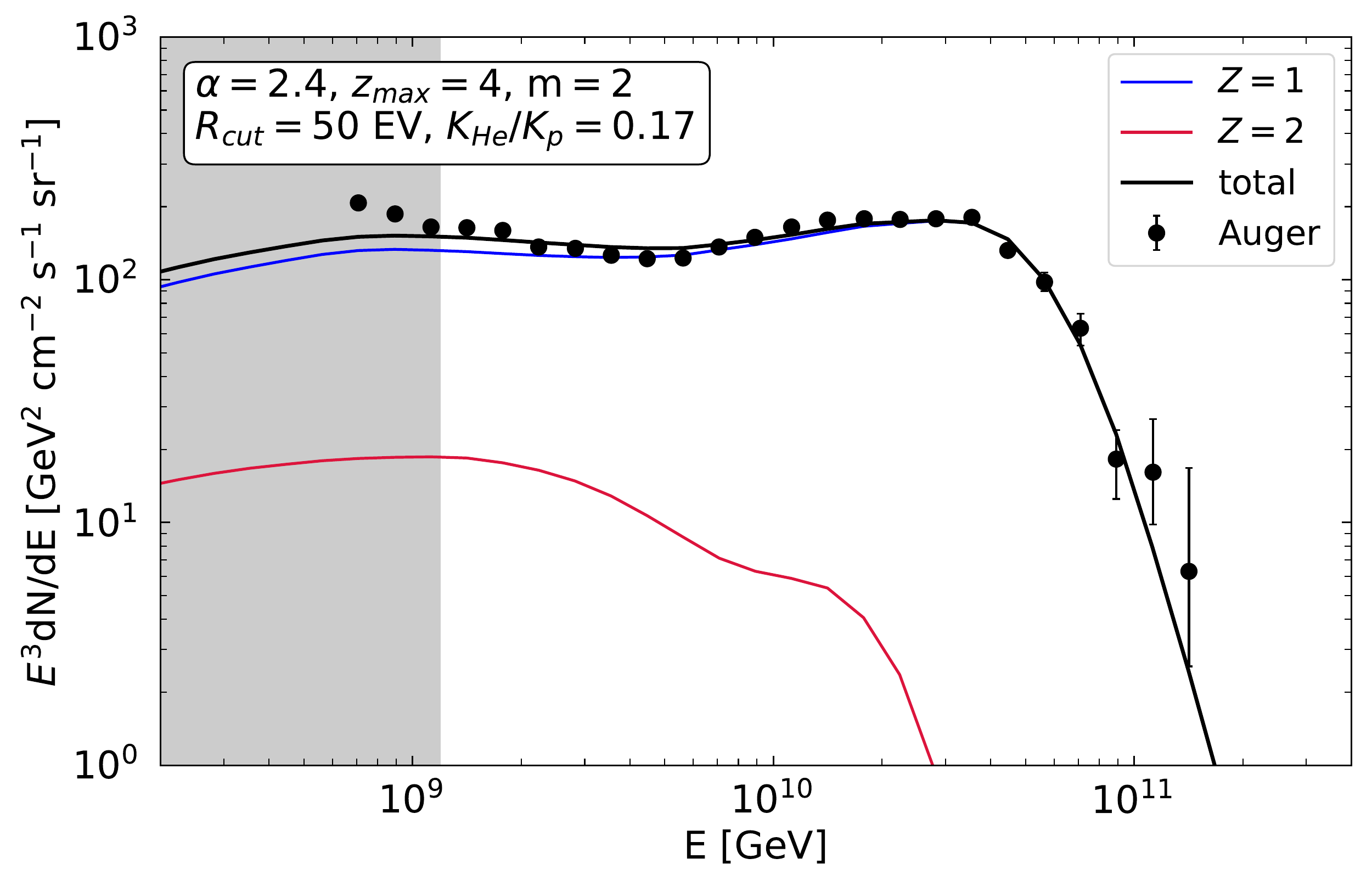}
\end{subfigure}%
\begin{subfigure}{.5\textwidth}
  \centering
  \includegraphics[width=8.0 cm, height = 5.2 cm]{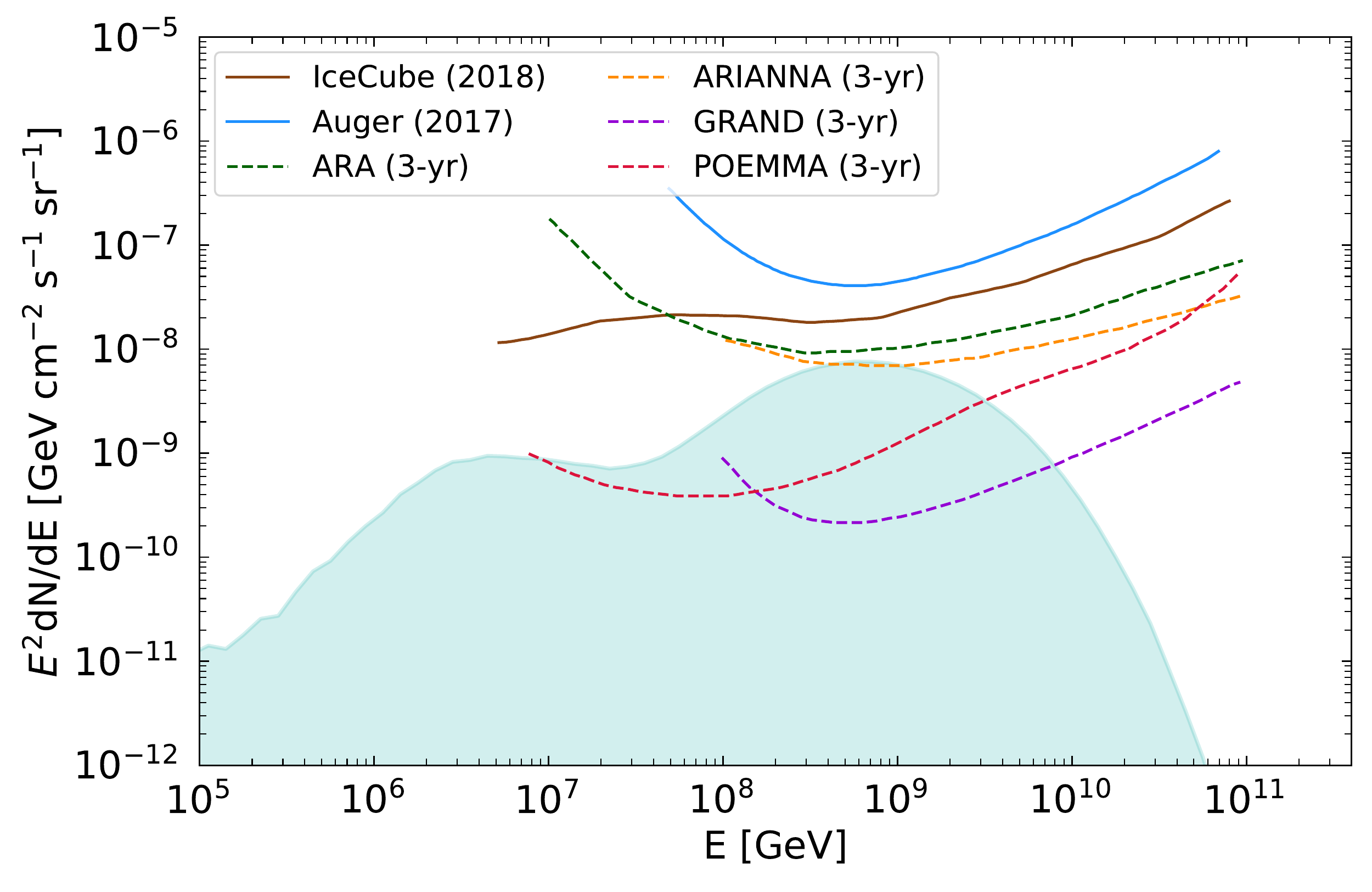}
\end{subfigure}
\caption{\small{UHECR spectra (left) and cosmogenic neutrino spectra (right) for $\alpha=2.4$. The top (case 13) and bottom (case 23) panels show the best-fit cases listed in Appendix A for which the difference in the cosmogenic neutrino flux is the maximum.}}
\label{fig:2.4}
\end{figure*}

\begin{figure*}[hbt]
\begin{subfigure}{.5\textwidth}
  \centering
  \includegraphics[width=8.0 cm, height = 5.2 cm]{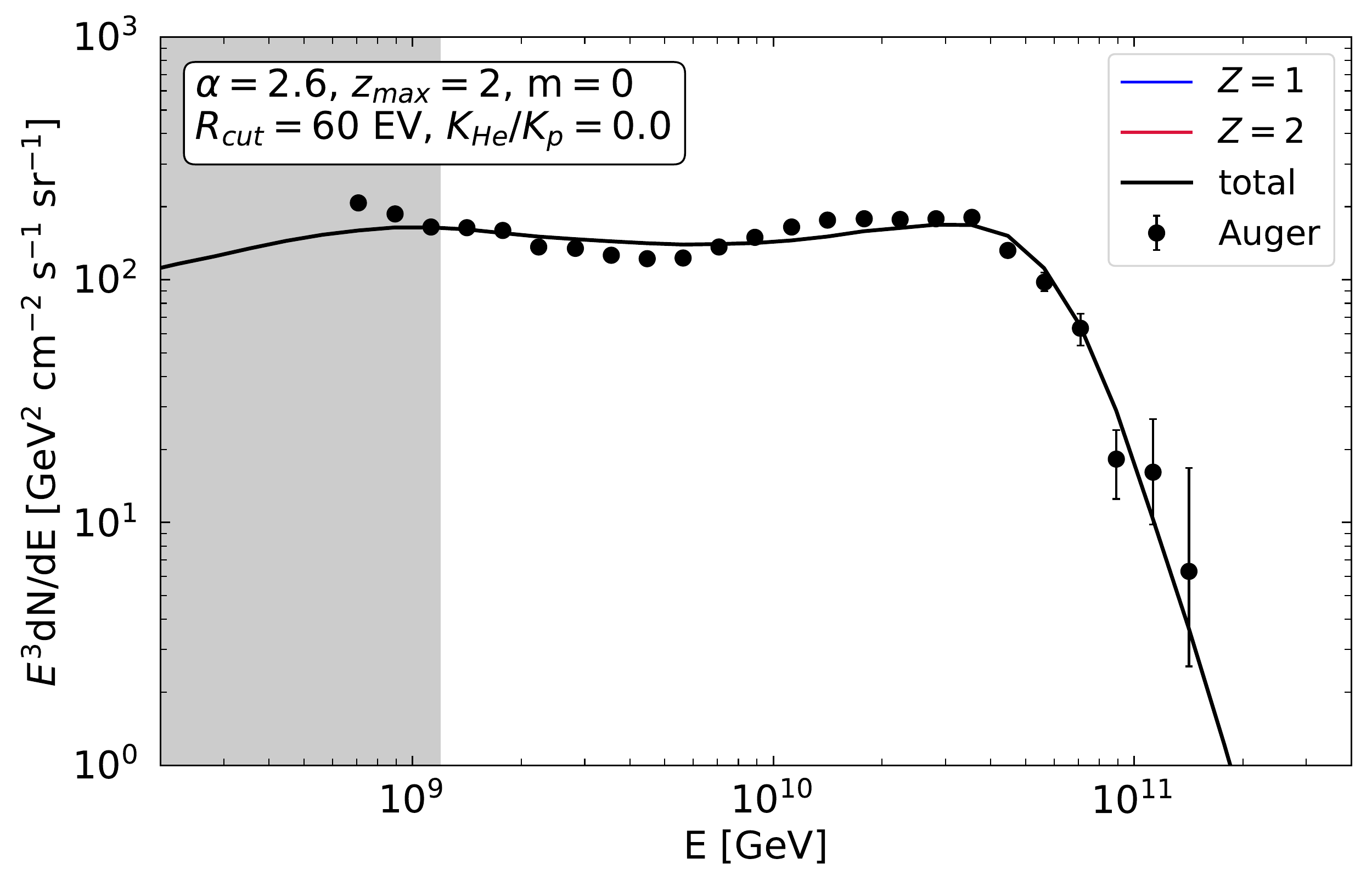}
\end{subfigure}%
\begin{subfigure}{.5\textwidth}
  \centering
  \includegraphics[width=8.0 cm, height = 5.2 cm]{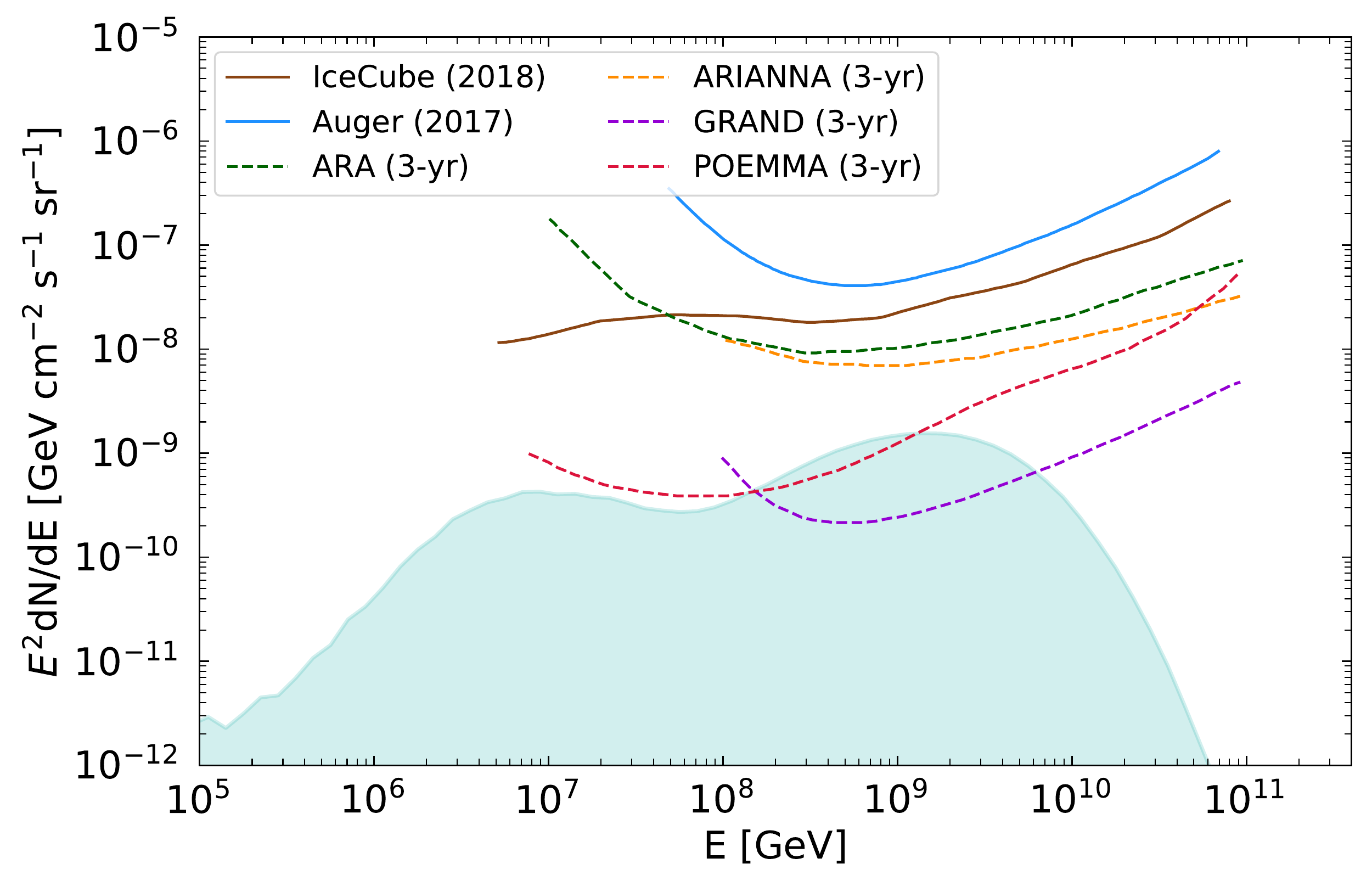}
\end{subfigure}
\begin{subfigure}{.5\textwidth}
  \centering
  \includegraphics[width=8.0 cm, height = 5.2 cm]{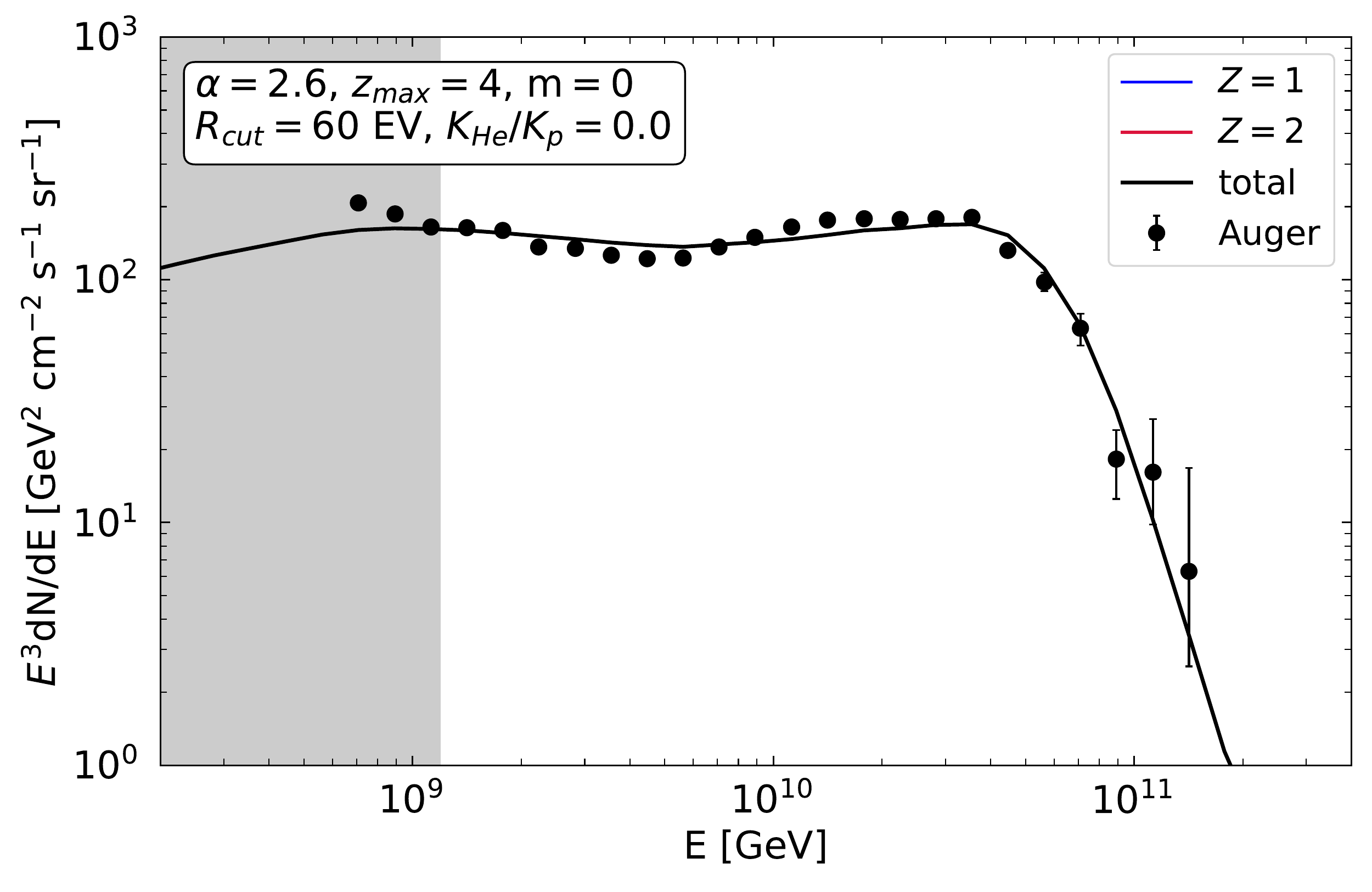}
\end{subfigure}%
\begin{subfigure}{.5\textwidth}
  \centering
  \includegraphics[width=8.0 cm, height = 5.2 cm]{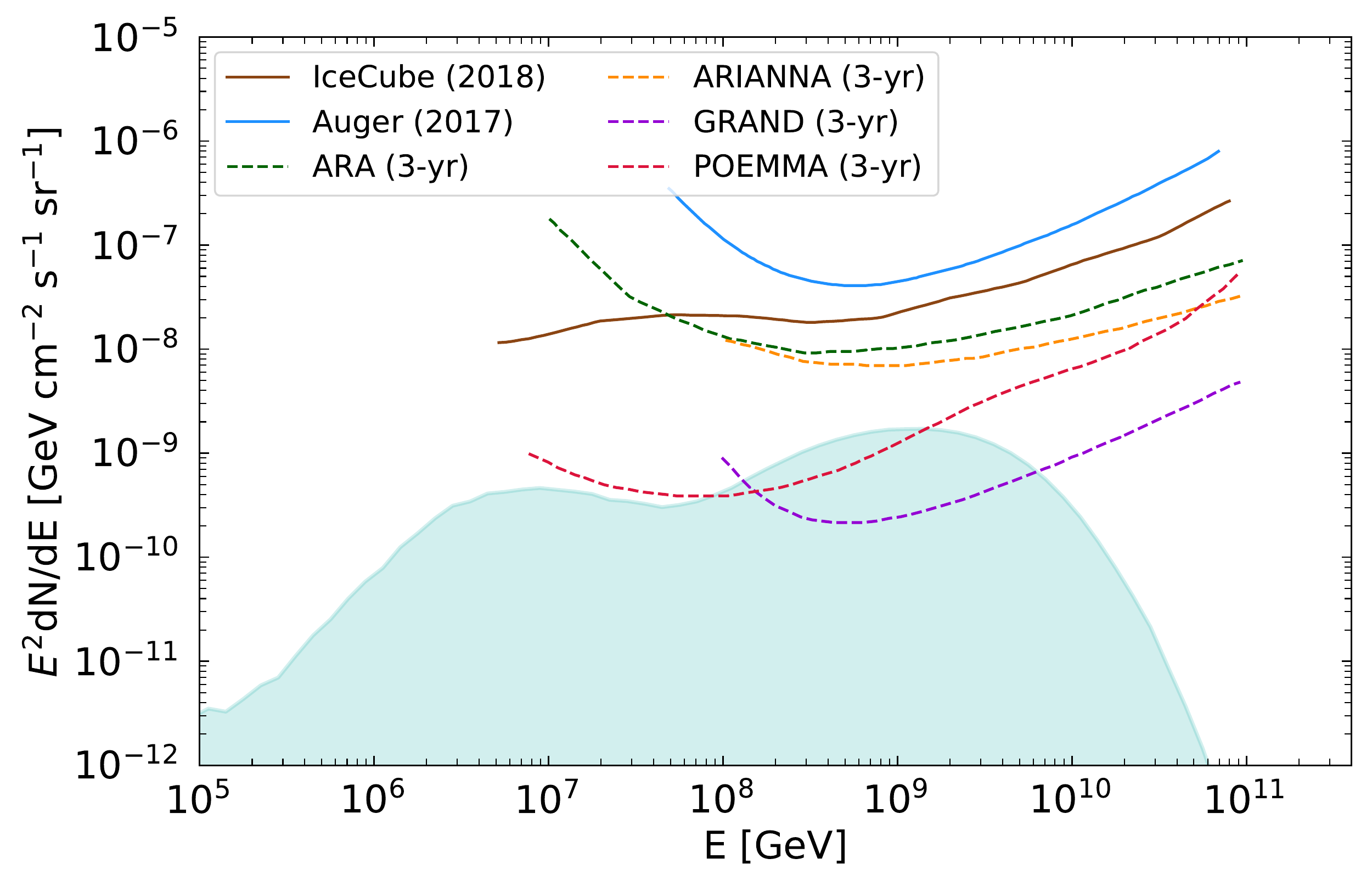}
\end{subfigure}
\caption{\small{UHECR spectra (left) and cosmogenic neutrino spectra (right) for $\alpha=2.6$. The top (case 25) and bottom (case 33) panels show the best-fit cases listed in Appendix A for which the difference in the cosmogenic neutrino flux is the maximum.}}
\label{fig:2.6}
\end{figure*}

\begin{figure*}[hbt]
\centering
\begin{subfigure}{.33\textwidth}
  \centering
  \includegraphics[width=6.0 cm, height = 3.9 cm]{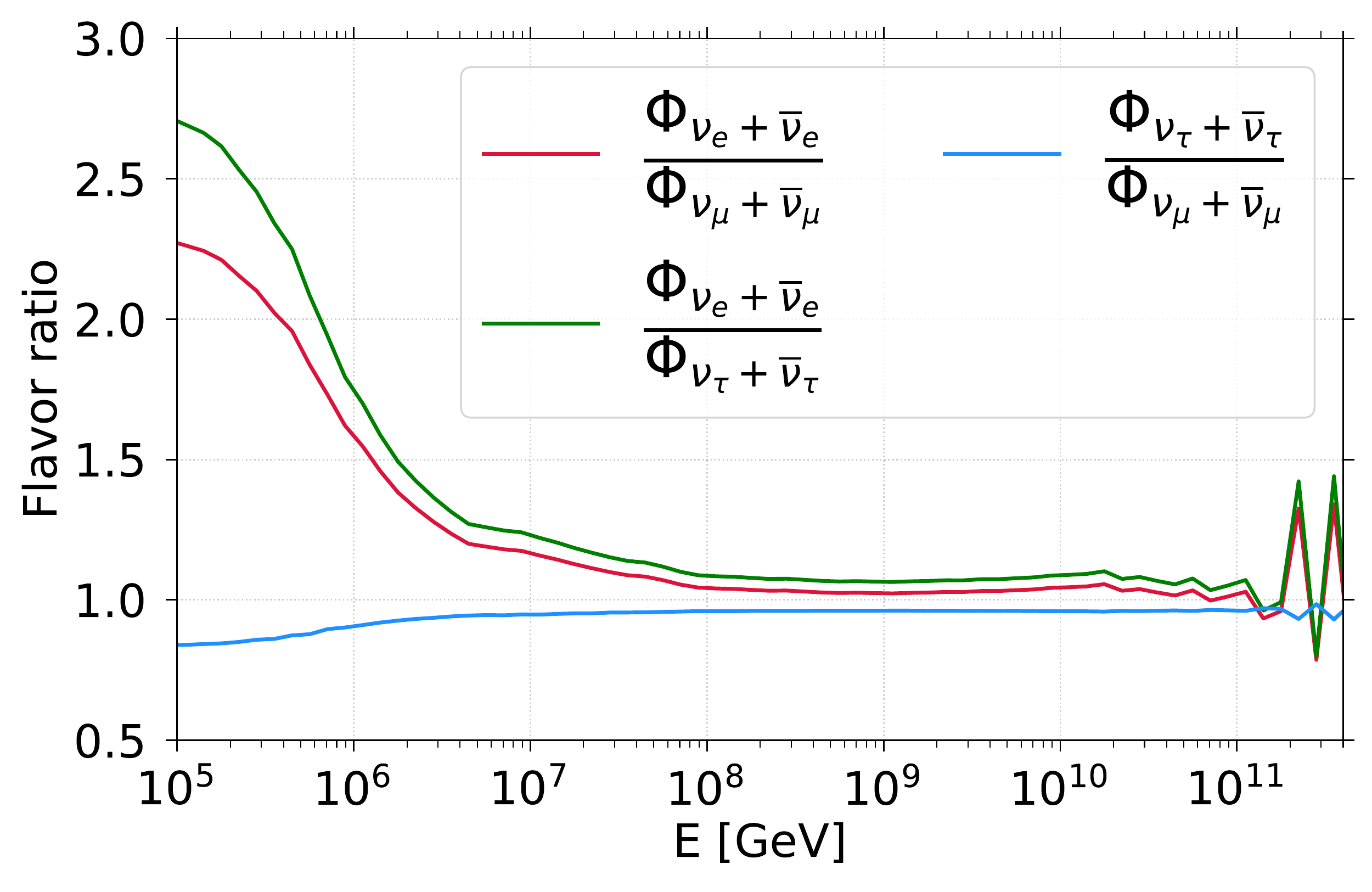}
\end{subfigure}%
\begin{subfigure}{.33\textwidth}
  \centering
  \includegraphics[width=6.0 cm, height = 3.9 cm]{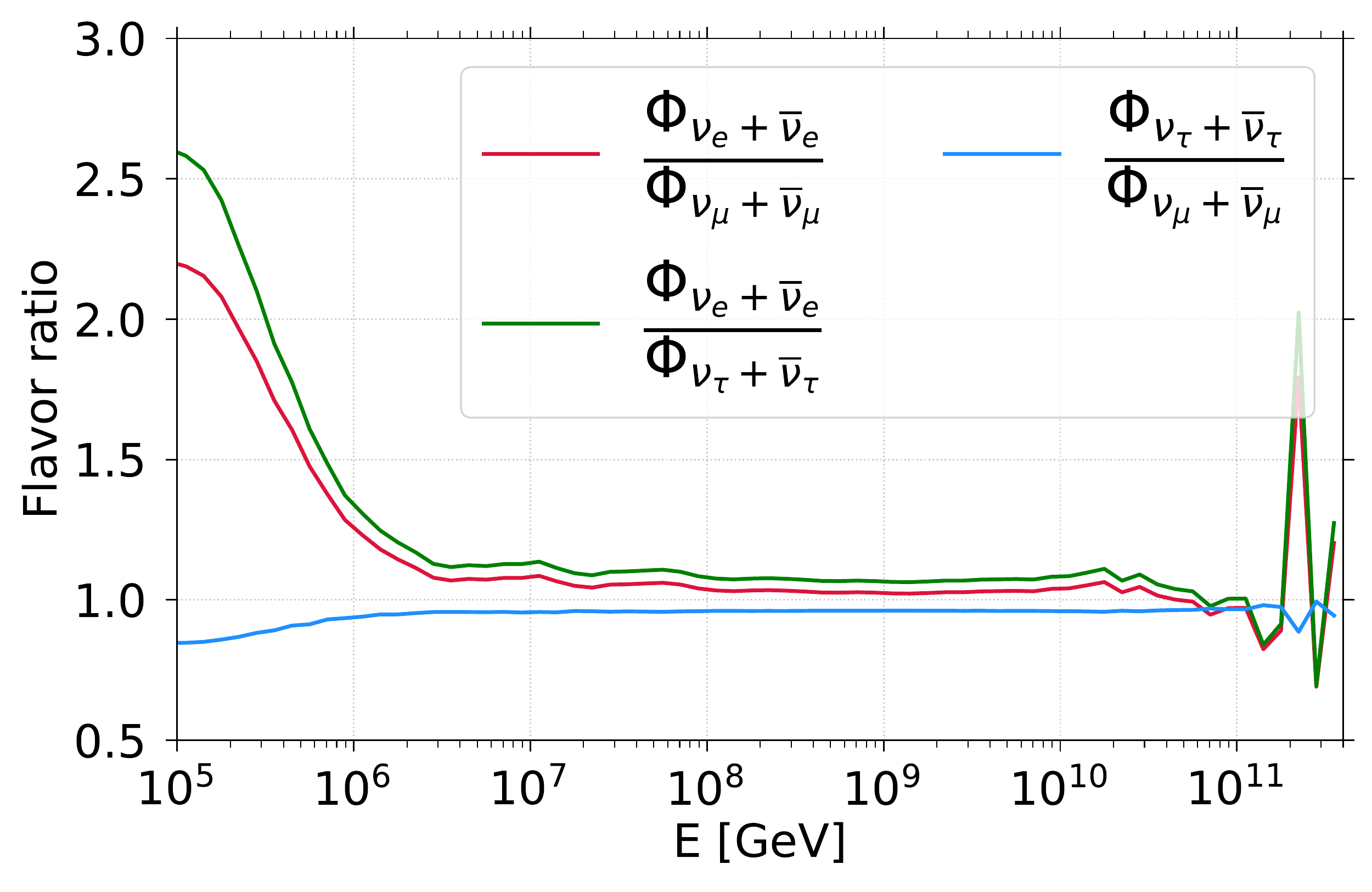}
\end{subfigure}%
\begin{subfigure}{.33\textwidth}
  \centering
  \includegraphics[width=6.0 cm, height = 3.9 cm]{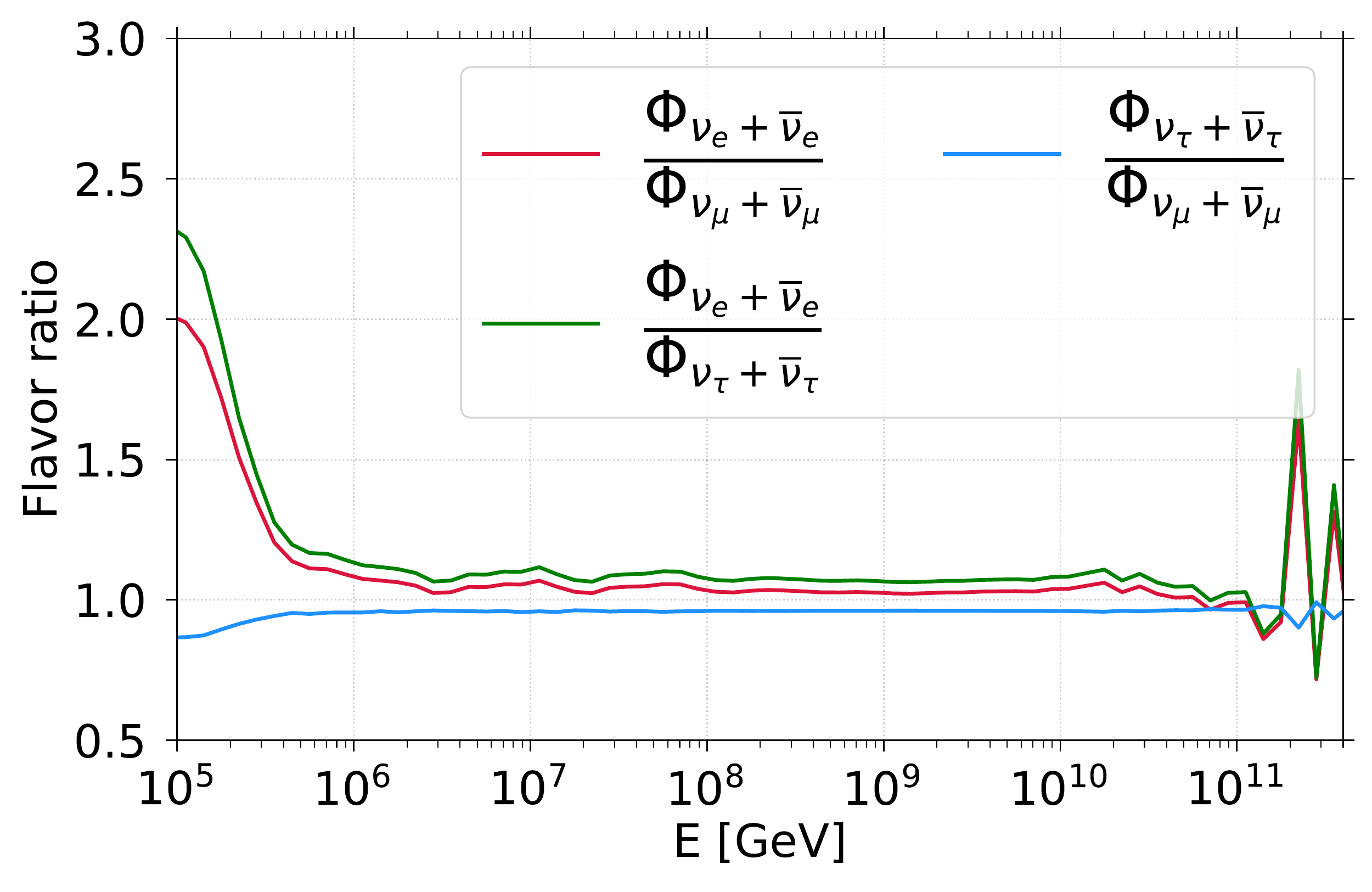}
\end{subfigure}

\begin{subfigure}{.33\textwidth}
  \centering
  \includegraphics[width=6.0 cm, height = 3.9 cm]{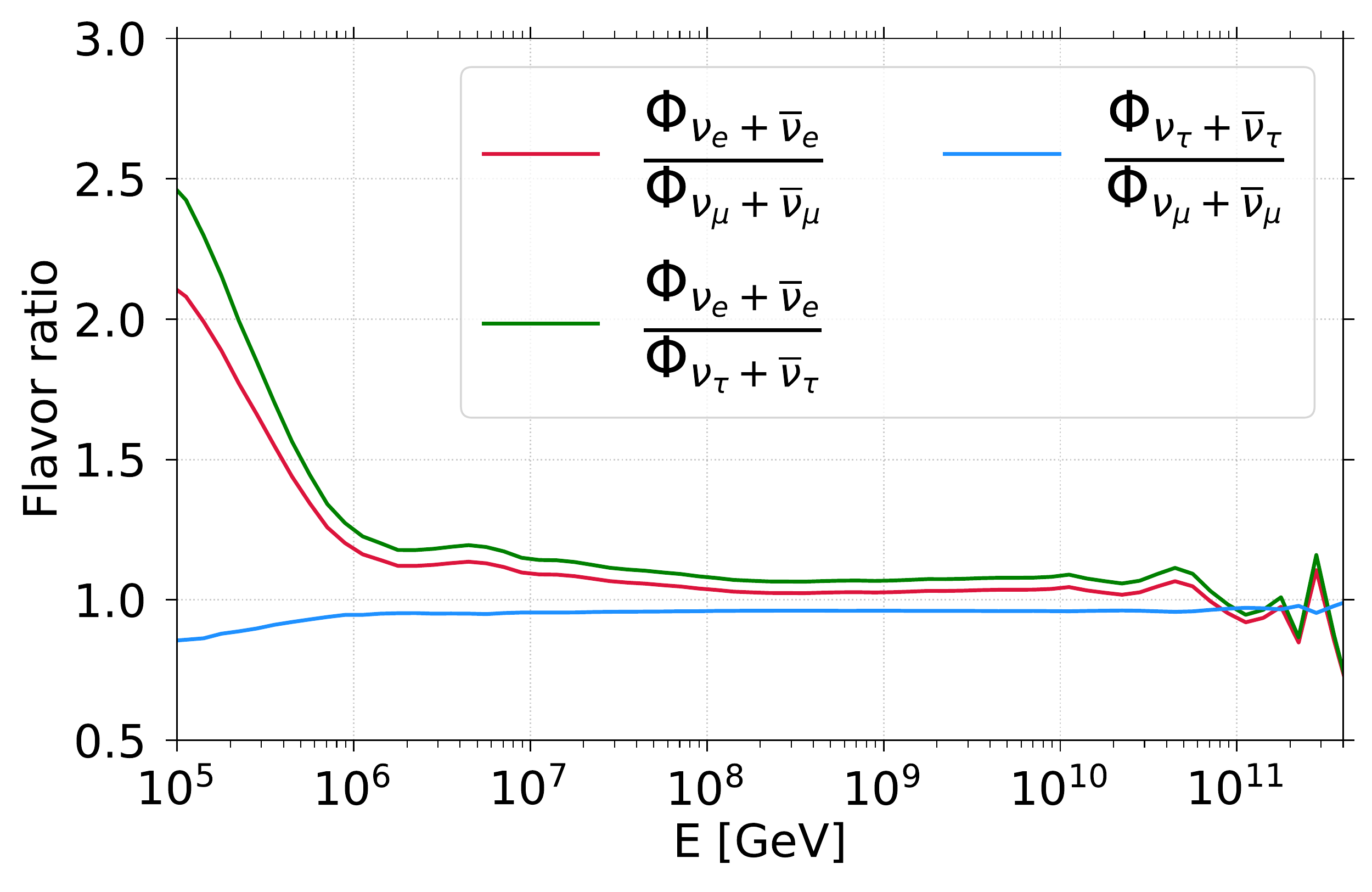}
\end{subfigure}%
\begin{subfigure}{.33\textwidth}
  \centering
  \includegraphics[width=6.0 cm, height = 3.9 cm]{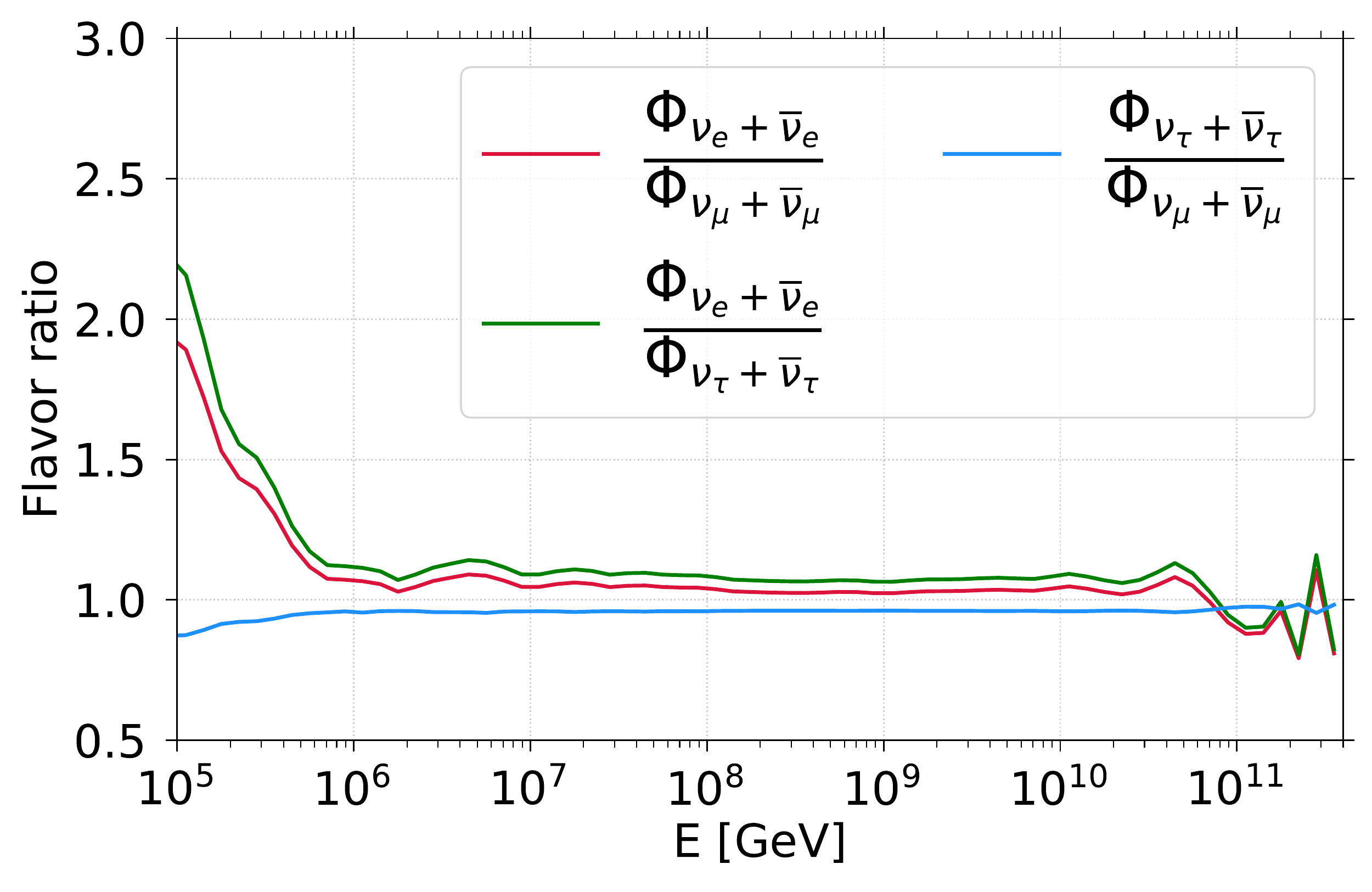}
\end{subfigure}%
\begin{subfigure}{.33\textwidth}
  \centering
  \includegraphics[width=6.0 cm, height = 3.9 cm]{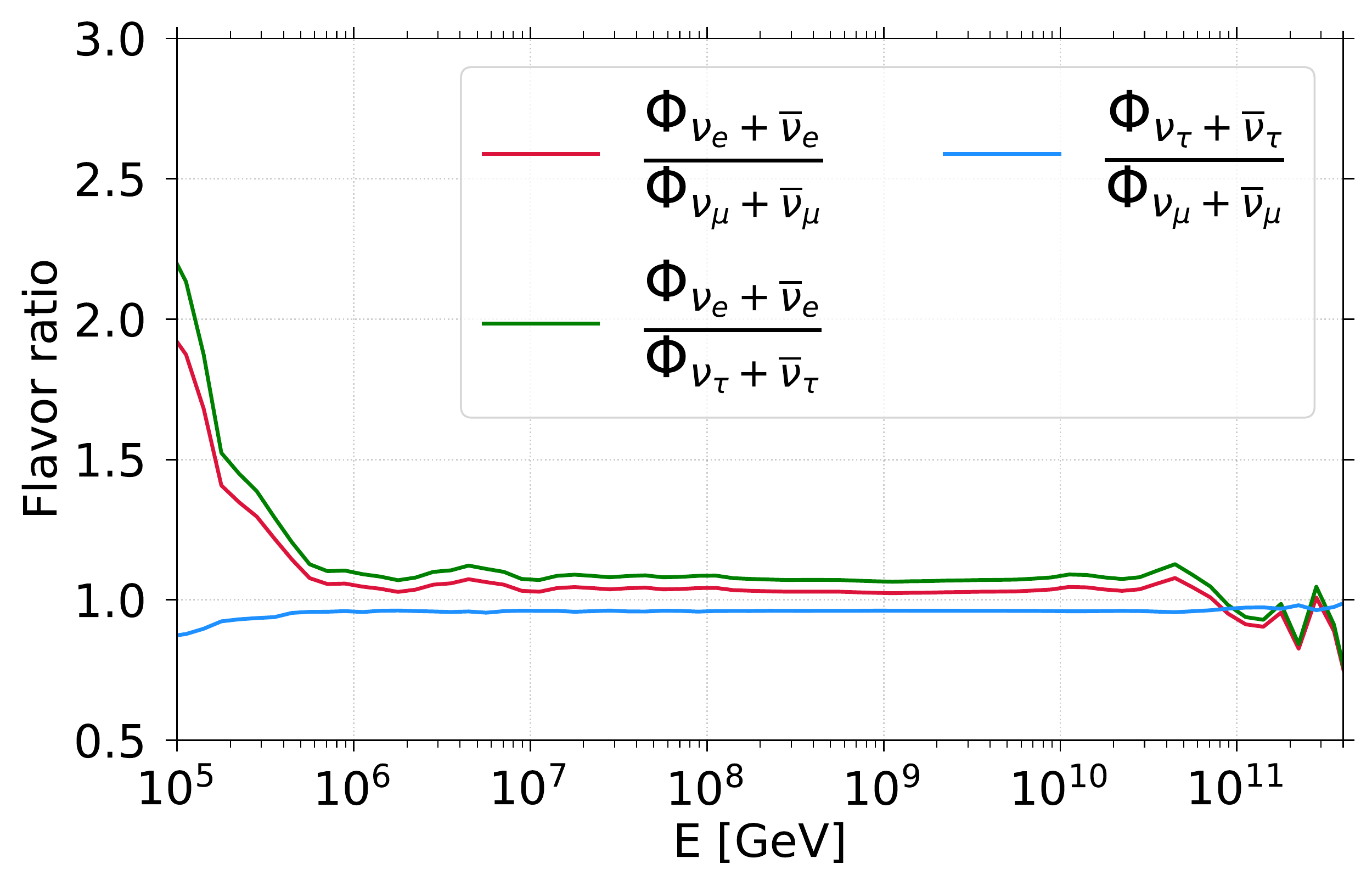}
\end{subfigure}
\caption{\small{Ratio of the neutrino flux of different flavors for the best fit cases. The left panels, middle panels and right panels are for $\alpha$ = 2.2, 2.4 and 2.6 respectively. The top and bottom figures correspond to the case with lowest neutrino flux and maximum neutrino flux respectively for each $\alpha$, i.e., for the cases plotted in Figs.~\ref{fig:2.2}--\ref{fig:2.6}}}
\label{fig:flavor}
\end{figure*}

We study the parameter space of UHECR sources and thus, of cosmogenic neutrino and photon fluxes for ``CTD" propagation model using two different astrophysical scenarios. In one case, we try to obtain a fit to the UHECR spectrum measured by PAO with only proton ($^1$H) and helium ($^4$He) as the primary composition at injection. In another case, we consider a mixed composition of four representative stable nuclei: hydrogen ($^1$H), helium ($^4$He), nitrogen ($^{14}$N) and silicon ($^{28}$Si). In the former case, it is possible to explain the UHECR spectrum over the entire ultrahigh energy range starting from $10^{18}$ eV. Whereas in the latter, a fit is possible for only the super-ankle region ($E>10^{18.7}$ eV). We find there is a marked difference in the feasible range of values for parameters like injection spectral index ($\alpha$), source evolution index ($m$), cutoff rigidity ($R_{\rm cut}$), etc., for the two frameworks. In the following, we do a comparative study of both scenarios. 

We begin our analysis by exploring the best-fits to UHECR spectrum, possible for p+He composition at injection. We constrain the allowed range of parameters from cosmogenic neutrino flux, as well as from composition measurement by PAO \citep{Aab17}. The parameters and their range of values studied for the p+He model are given in Table \ref{tab:param}. The cutoff rigidity $R_{\text{cut}}$ of injected primaries is varied between $40-100$ EV in steps of 10 EV. The source evolution index $m$ is varied through 0, 1, 2, 3. The sources are restricted to $z_{\text{min}} \leqslant z \leqslant z_{\text{max}}$, where the minimum source redshift is fixed at $z_{\text{min}} = 0.0007$, corresponding to the distance to Centaurus A (the nearest AGN). Since the star formation rate peaks near $z\simeq 2$, $z_{\text{max}}$ is varied through 2, 3 and 4. The maximum redshift in the simulations are thus well above $z=0.06$, the GZK horizon. We investigate three particular cases of source spectral index at injection, viz. $\alpha = 2.2$, $2.4$, $2.6$. The abundance fractions, $K_{\rm p}$ and $K_{\rm He}$ is varied from $0$\% to $100$\% with a precision of $0.1$\%, restricted by the condition, $K_{\rm p}+K_{\rm He}=100$\%. For each possible combination of \{$\alpha$, $z_{\text{max}}$\}, we vary $m$ and calculate the best-fit value of $R_{\rm cut}$ and composition. The results of the scan is given in Table~\ref{tab:bestfit} of Appendix~\ref{app:best}.

The fit of the simulated spectrum to Auger data is done for $E>10^{18}$ eV, with the 21 highest energy data points \citep{Aab17}. The goodness-of-fit to the spectrum is computed using a standard $\chi^2$ analysis,
\begin{equation}
\chi_{\rm spec}^2 = \sum_{i=1}^{N} \bigg[\dfrac{y_i^{\rm obs}(E) - y_i^{\rm mod}(E;a_M)}{\sigma_i}\bigg]^2
\end{equation} 
Here, $y_i^{\rm obs}(E)$ is the observed value of the UHECR flux and $y_i^{\rm mod}(E;a_M)$ are the simulated values, at specific energies, respectively, and $a_M$ are the values of $M$ parameters in the simulation. The standard error of each observed value is given by $\sigma_i$. We take $\sigma_i = \sqrt{y_i^{\rm obs}(E)}$, where no data for errors are given. We add asymmetric errors in quadrature to calculate $\chi^2$ values. The $\chi^2$ value for each of the best-fit cases is given in Table~\ref{tab:bestfit}. The number of degrees of freedom (d.o.f.), $N_{\rm d}=21-3-1=17$, since we fit the simulated spectrum to 21 Auger data points and vary three parameters $R_{\rm cut}$, $m$, $K_{\rm p}$ for fixed \{$\alpha$, $z_{\text{max}}$\}. We consider the normalization factor to be an additional free parameter. Throughout the study, we restrict ourselves to best-fit cases with $\chi^2<27.95$, i.e., within $2\sigma$ standard deviations for 17 d.o.f. 

We present some of the best-fit spectra found for p+He model in Subsec.~\ref{subsec:uhecr}, along with the corresponding cosmogenic neutrino fluxes. The results deduced from neutrino flux calculation are elaborated in Subsec.~\ref{subsec:neutrino}. We present expected ratios of neutrino fluxes of different flavors in Subsec.~\ref{subsec:neu_comp}, which can serve as discriminators between different composition models. A finer variation of $R_{\text{cut}}$ and $m$ is done afterwards in Subsec.~\ref{subsec:correlation}. There we study the correlation between fit parameters and explain the allowed range of their values in more detail. We demonstrate in Sebsec.~\ref{subsec:photons} that cosmogenic photon flux can constrain mass composition in certain cases. In Subsec.~\ref{subsec:heavy}, we compare our results for p+He model with that of p+He+N+Si model at injection.

\subsection{\label{subsec:uhecr}Fits to the UHECR spectrum\protect}

In Table~\ref{tab:bestfit} we list all the 36 best-fit cases obtained from the parameter scan. Restricting the best-fits to $\chi^2<27.95$ immediately disfavors the $m=0$ cases for $\alpha=2.2$ and $m \neq 0$ cases for $\alpha=2.6$. Also, we discard the $m=3$ cases for $\alpha=2.4$, as it corresponds to almost pure proton at injection and does not comply with composition measurements by Auger. Only $m=0$ choice gives acceptable $\chi^2$ for $\alpha = 2.6$, although for all $\alpha = 2.6$ cases the spectrum is composed of pure proton and hence  are disfavored. In the left panels of Figs.~\ref{fig:2.2}, \ref{fig:2.4} and \ref{fig:2.6} we show some of our fits to the UHECR spectrum in $E^3(dN/dE)$ units for $\alpha=$ 2.2, 2.4 and 2.6, respectively. The right panels show the corresponding cosmogenic neutrino flux for those fits. For each $\alpha$, we choose the two best-fit UHECR scenarios (top and bottom panels) from the allowed cases, for which the difference in the cosmogenic neutrino fluxes is maximum. The parameter set for the maximum allowed neutrino flux coincides with that for the minimum $\chi^2$ in all $\alpha$ values considered. This corresponds to cases - 2 and 12 from top to bottom in Fig.~\ref{fig:2.2}; cases - 13 and 23 in Fig.~\ref{fig:2.4}; cases - 25 and 33 in Fig.~\ref{fig:2.6}, from the allowed cases. The shaded region in the plots indicates the part excluded in calculating the $\chi^2$ values. The highest energy Auger data points beyond 40 EeV are well covered for our chosen range of $R_{\text{cut}}$. 

We make the following observations from our study of the model parameters fitting data:
\begin{itemize} 
\item The required helium to proton fraction $K_{\rm He}/K_{\rm p}$ decreases with increasing value of $\alpha$ and essentially no He is required at injection for $\alpha = 2.6$ (see Fig.~\ref{fig:2.6}). 
\item The He flux falls off sharply beyond a few EeV due to increased photodisintegration on EBL, conforming with the predictions by Gerasimova and Rozental \citep{gerasimova}. This elucidates the proton dominance at the highest energies, making GZK cutoff a conspicuous phenomenon.
\item For $\alpha = 2.2$, the fraction $K_{\rm He}/K_{\rm p} > 1$ and the He flux is comparable or dominating over the proton flux at $\lesssim 1$~EeV.  As such, a changing composition contributes to the ankle feature. For $\alpha = 2.6$ the ankle is purely due to $\mathrm{e^+}\mathrm{e^-}$ pair production with the CMB photons \cite{Berezinsky06}.  
\item For $\alpha=2.6$, only a uniform source distribution, i.e., $m=0$ is able to fit the data. With higher values of $m$, the $\chi^2$ value increases rapidly, indicating the fit worsens. 
\item Increasing $m$ causes hardening of the spectrum in the energy region below the ankle. For a particular $\alpha$ and $z_{\text{max}}$, the fit improves on increasing $m$ and thereby lowering $K_{\rm He}/K_{\rm p}$, implying a lower He abundance at the sources for higher $m$ values.
\item There is no significant change in the UHECR spectral models due to the variation of $z_{\text{max}}$ beyond a redshift of 2.0.
\end{itemize}
However, the $\chi^2$ values suggest a better fit for lower $\alpha$ values among the allowed cases. But lowering $\alpha$ below 2.2 makes it difficult to obtain a fit from $\approx$ 1 EeV with only proton and helium at injection. The fluctuation in shower depth distribution data indicates a mass composition between H and He up to $10^{19.5}$ eV. Hence, the pure proton composition obtained in $\alpha=2.6$ cases is disfavored. This explains our choice of spectral indices in the range $2.2\leqslant \alpha \leqslant 2.6$.

\subsection{\label{subsec:neutrino}Cosmogenic neutrino fluxes\protect}

We calculate the neutrino flux with CRPropa by taking into account all possible production channels and by using the same normalization factor used for fitting UHECR data in different cases (see table in Appendix A). We present the whole range of cosmogenic neutrino flux summed over all flavors, in  $E^2(dN/dE)$ units, possible within the p+He model in the right panels of Figs.~\ref{fig:2.2}--\ref{fig:2.6}. A double peak shape is a common feature to all neutrino spectra. The higher-energy bump at around $\approx 10^{18}$ eV is due to decay of pions produced in interactions of UHECRs with the CMB photons. The lower-energy bump at $\sim 10^{16}$ eV is due to a combination of neutron beta decay and decay of pions produced in interactions of UHECRs with the EBL photons. The flux values at the higher-energy peak for different cases are listed in the table in Appendix A. The main results from our study of all flavor cosmogenic neutrino fluxes are below.
\begin{itemize}
\item Although no significant change in UHECR spectrum is seen on variation of $z_{\text{max}}$ beyond a redshift of 2.0, the cosmogenic neutrino flux on the other hand increases with increasing $z_{\text{max}}$, keeping all other parameters fixed.
\item The flux at the higher-energy peak is generally higher than the lower energy peak for a harder ($\alpha = 2.2, 2.4$) injection spectrum. The lower-energy peak becomes more pronounced for a softer ($\alpha = 2.6$) injection spectrum. The flux ratio between the two peaks reaches up to an order of magnitude, for a harder injection spectrum.
\item The exact position of the peaks and the flux values depend on the maximum distance and redshift evolution of sources as well as the relative abundance of proton and helium.
\end{itemize} 

The neutrino fluxes in Figs.~\ref{fig:2.2}--\ref{fig:2.6} are compared to the current and upcoming detector sensitivities as well. We show the detection sensitivity curves for Auger \citep{augerneu1, augerneu2} and the flux upper limits from IceCube \citep{aartsen15, aartsen16, aartsen18} along with the extrapolated 3-year sensitivities for the proposed detectors ARIANNA \citep{arianna}, ARA \citep{ara}, POEMMA \citep{poemma1, poemma2} and GRAND \citep{grand1, grand2, moller18}. The upcoming Mediterranean detector KM3NeT \cite{Adrian-Martinez:2016fdl} and the proposed extension of the IceCube detector called IceCube-Gen2 \cite{vanSanten:2017chb} can also probe cosmogenic neutrino fluxes in near term. 

The PAO is effective at searching for neutrinos of energies exceeding 0.1 EeV by selecting inclined showers that have significant electromagnetic component. The range of neutrino fluxes obtained in our simulations are clearly below the differential upper limit $E^2\Phi_{\nu} \approx 4\times 10^{-8}$ GeV cm$^{-2}$ s$^{-1}$ sr$^{-1}$ imposed by PAO at 0.6 EeV. We multiply the single-flavor neutrino flux limit of Auger by a factor of 3 to obtain the all-flavor neutrino flux limit, assuming an equal flavor ratio.

The 90\% C.L. all-flavor differential flux upper limit from 9-years of IceCube data sample based on extreme high energy (EHE) neutrino events above $5\times10^{6}$ GeV is shown in solid brown line. Two EHE events were observed in the 9-yr analysis, which is compatible with a generic astrophysical origin and inconsistent with the cosmogenic hypothesis (for details, see \citep{aartsen18}). In our calculations, the limit by IceCube just touches the CMB peak of the case 12 corresponding to maximum neutrino flux for $\alpha=$ 2.2. This indicates a detection should be possible in the near future, with a further increase in exposure time. For the pure proton injection model ($\alpha=2.6$), the cosmogenic neutrino fluxes are too low to be detected by IceCube in the future. 

The sensitivities for the upcoming detectors are calculated from the simulation of antenna response. ARA and ARIANNA, proposed to be built in Antarctica, aims at using Askaryan effect to detect interactions of the cosmogenic neutrinos above 1 EeV with ice. With comparable 3-yr sensitivities, both detectors would be able to probe few of our harder ($\alpha = 2.2, 2.4$) injection spectrum cases (e.g., cases 12, 23). The sensitivities of POEMMA and GRAND are expected to be much better and would be able to probe cosmogenic fluxes for all the cases we explored. In particular, GRAND plans to reach an all-flavor integral limit of $\sim 1.5 \times 10^{-10}$ GeV cm$^{-2}$ s$^{-1}$ sr$^{-1}$ above $5 \times 10^{17}$ eV and a subdegree angular resolution \citep{grand2}. For a neutrino flux of $10^{-8}$ GeV cm$^{-2}$ s$^{-1}$ sr$^{-1}$, the GRAND sensitivity corresponds to a detection of $\sim100$ events after three years of observation. The maximum neutrino flux obtained in our study is $2.079\times10^{-8}$ GeV cm$^{-2}$ s$^{-1}$ sr$^{-1}$ corresponding to $\alpha=2.2$, $z_{\rm max}=4$, $m=3$ (case 12). This implies that GRAND will either detect cosmogenic neutrinos or constrain these model parameters from a few years of observation. The most pessimistic scenario predicts a neutrino flux $2.347\times10^{-9}$ GeV cm$^{-2}$ s$^{-1}$ sr$^{-1}$, obtained for $\alpha=2.2$, $z_{\rm max}=2$, $m=1$ (case 2). In that case, the probability of detection is low $\sim 20$ events in 3 years, nearly one-tenth of the event rate for the most optimistic scenario.

\subsection{\label{subsec:neu_comp}Cosmogenic neutrino flux components\protect}

After propagation over astrophysical distances, the probability of neutrino flavor conversion from $\nu_\alpha$ to $\nu_\beta$ is given by $P_{\alpha\beta} = \sum_{j=1}^{3} \left| U_{\beta j} \right|^2 \cdot \left| U_{\alpha j} \right|^2$, where $\alpha ,\beta = e, \mu, \tau$ and $U$ is the Pontecorvo-Maki-Nakagawa-Sakata (PMNS) mixing matrix between the neutrino flavor and mass eigenstates.  We use the current best-fit values of the mixing angles (for the normal mass hierarchy): $\sin^2\theta_{12} = 0.297$, $\sin^2\theta_{23} = 0.425$, $\sin^2\theta_{13} = 0.0215$ and the CP-violating phase $\delta = 1.38\pi$ \cite{Capozzi:2017ipn}.  The corresponding probability matrix is
%
\begin{eqnarray}
\left(
\begin{array}{ccc}
 P_{ee}     & P_{e\mu}    &  P_{e\tau} \\
 P_{\mu e}  & P_{\mu\mu}  &  P_{\mu\tau} \\
 P_{\tau e} & P_{\tau\mu} &  P_{\tau\tau} 
\end{array}
\right) 
\approx 
\left(
\begin{array}{ccc}
 0.56 & 0.24 & 0.20 \\
 ...  & 0.38 & 0.38 \\
 ...  & ...  & 0.42 
\end{array}
\right)
\end{eqnarray}
%
which is the same for neutrinos and antineutrinos with $P_{\alpha\beta} = P_{\beta\alpha}$.  Interestingly, for $\delta = 0$, there is a $\sim 10\%$ change in the probabilities: $P_{e\mu} \approx 0.28$, $P_{e\tau}\approx 0.17$, $P_{\mu\mu}\approx 0.35$ and $P_{\tau\tau}\approx 0.46$.  The probabilities $P_{ee}$ and $P_{\mu\tau}$ remain unchanged.  In principle, $\delta$ can be probed with precise knowledge of the mixing angles and cosmogenic fluxes. 

We calculate the cosmogenic neutrino fluxes of different flavors on the Earth from the fluxes generated by the CRPropa code $\Phi^0_\alpha$ as
\begin{eqnarray}
\Phi_{\nu_\alpha + {\bar\nu}_\alpha} &=& 
P_{e\alpha} (\Phi^0_{\nu_e} + \Phi^0_{{\bar\nu}_e}) +
P_{\mu\alpha} (\Phi^0_{\nu_\mu} + \Phi^0_{{\bar\nu}_\mu}) \nonumber \\ && +
P_{\tau\alpha} (\Phi^0_{\nu_\tau} + \Phi^0_{{\bar\nu}_\tau})
\end{eqnarray}

The main discriminator for neutrino flavors at the ice/water Cherenkov detectors is event topology, namely tracks for $\nu_\mu$ charged-current events and showers for $\nu_e$ and $\nu_\tau$ charged-current events and for all neutral-current events.  At $\gtrsim 1$ PeV range, it is however, possible to discriminate $\nu_\tau$ events \cite{Learned:1994wg, Beacom:2003nh, DeYoung:2006fg} and flavor identification for all charged-current events could be possible.  In such a case the ratios of cosmogenic fluxes of different flavors can be written as
\begin{equation}
r_{\alpha/\beta} =\frac{\Phi_{\nu_\alpha + {\bar\nu}_\alpha}}
{\Phi_{\nu_\beta + {\bar\nu}_\beta}}
\end{equation}
For typical $1:2:0$ initial flavor ratios, the expected ratio on the Earth is just ratio of the probabilities given as
\begin{equation}
r_{\alpha/\beta} =\frac{P_{e\alpha} + 2P_{\mu\alpha}}{P_{e\beta} + 2P_{\mu\beta}}
\end{equation}

In Fig.~\ref{fig:flavor}, we plot the ratios $r_{e/\mu}$, $r_{\tau/\mu}$ and $r_{e/\tau}$ obtained from CRPropa simulations. The left, middle and right panels show the ratio of component fluxes for $\alpha=$ 2.2, 2.4 and 2.6, respectively. Top and bottom plots in Fig.~\ref{fig:flavor} represent the same cases as shown accordingly in top and bottom plots of Figs.~\ref{fig:2.2}--\ref{fig:2.6}. At the highest energy end, the ratios are not well defined due to few particles involved in the simulations. Below $\approx 10^{20}$~eV, the ratios are roughly constant for a number of decades in energy depending on different cases. These constant ratio parts are roughly consistent with the expected values of $r_{e/\mu} = 1.03$ (red lines), $r_{\tau/\mu} = 0.96$ (blue lines) and $r_{e/\tau} = 1.08$ (green lines); from typical pion-decay flavor ratios $1:2:0$ at production. A shift from these values at low energies is due to neutron beta decays and is an indicator of He/p ratio of the UHECR flux at injection. For example, in $\alpha = 2.2$ cases requiring larger He/p ratio, the deviation from constant flavor ratios happen at energies $\lesssim 10^{17}$~eV, while for the pure protons injection cases ($\alpha = 2.6$), the flavor ratios are constant down to $10^{15}$~eV.

\subsection{\label{subsec:correlation}Correlation between fit parameters\protect}

\begin{figure}
\centering
\includegraphics[width = 0.48\textwidth]{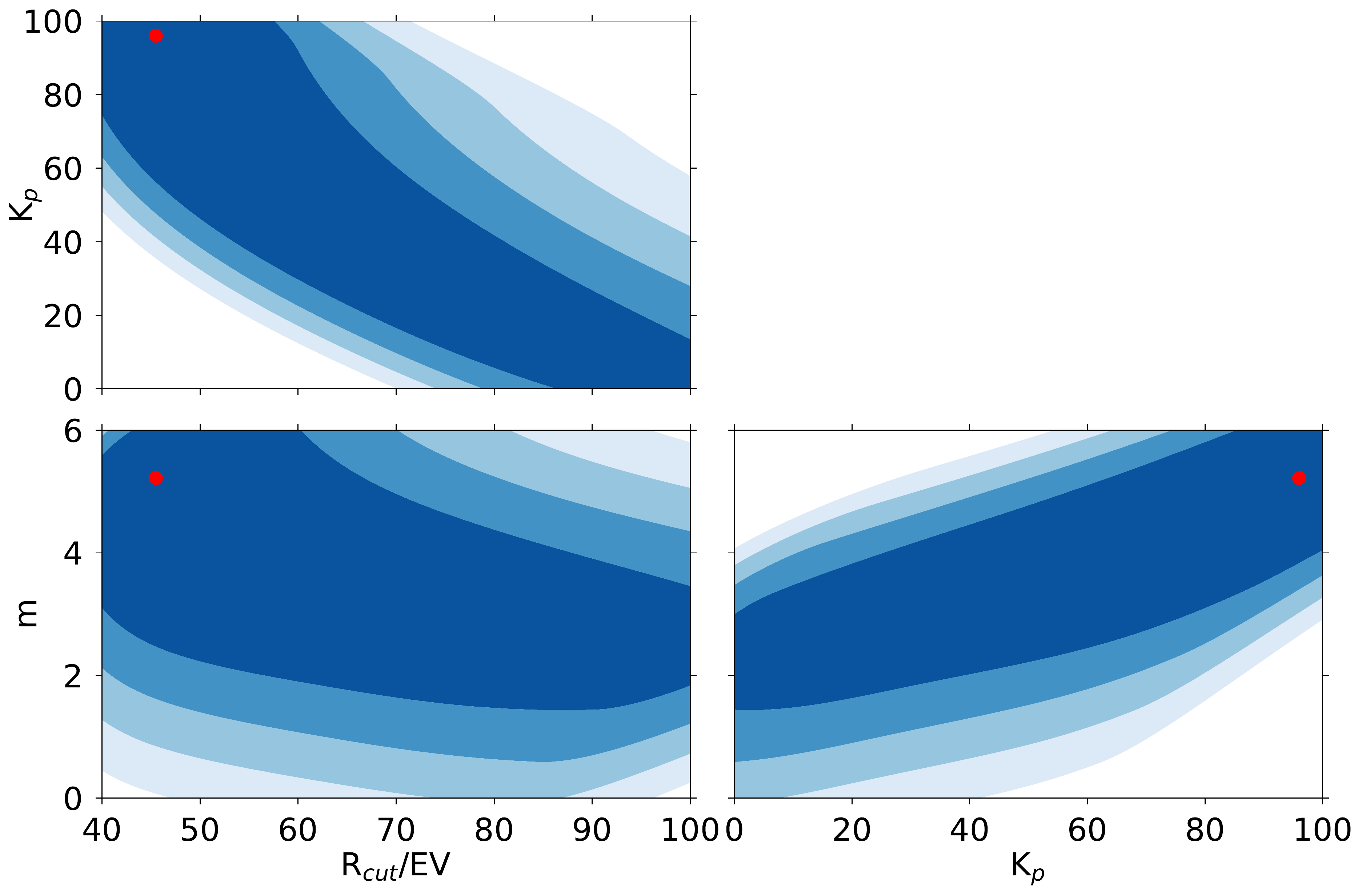}
\includegraphics[width = 0.48\textwidth]{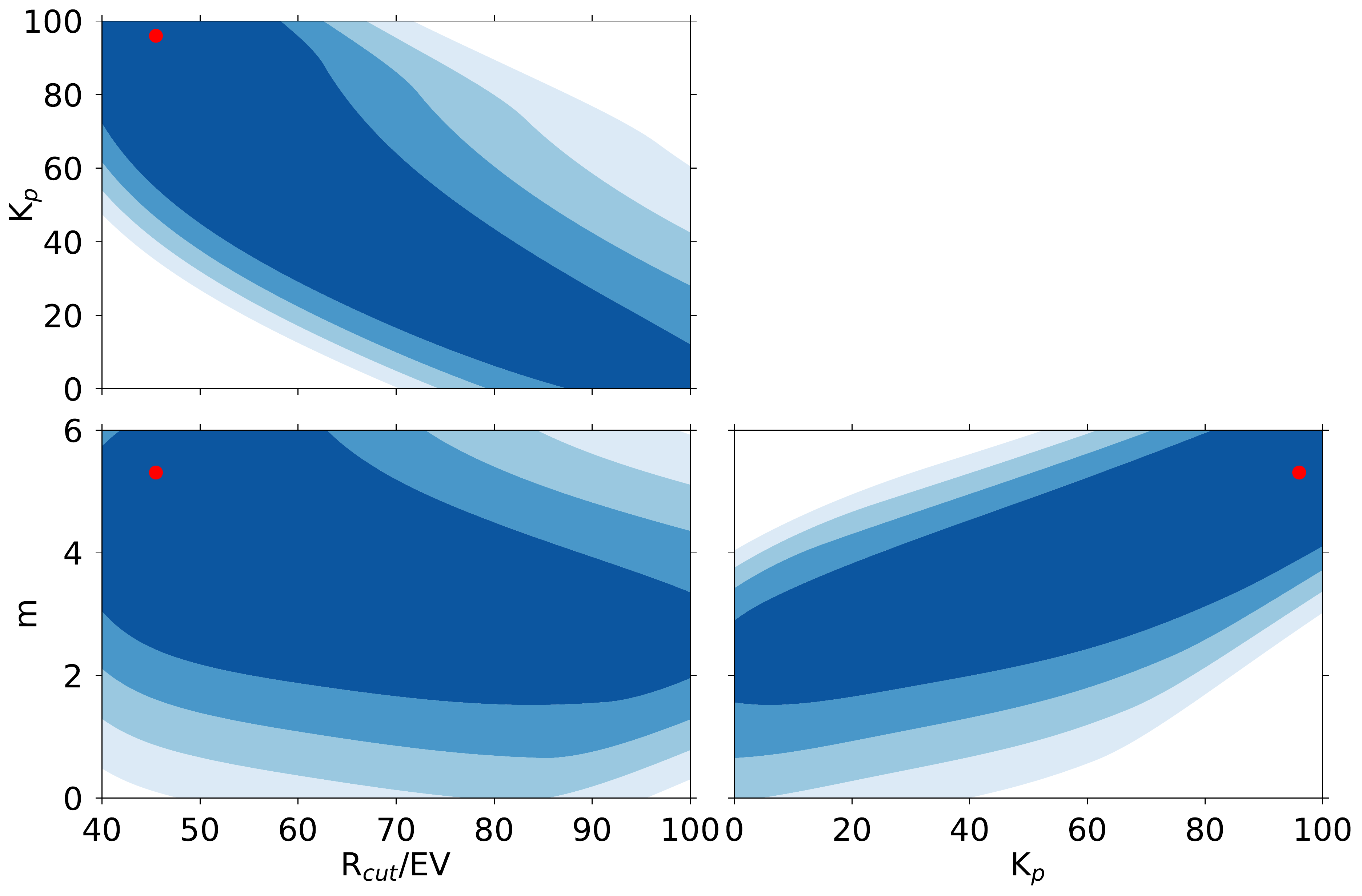}
\includegraphics[width = 0.48\textwidth]{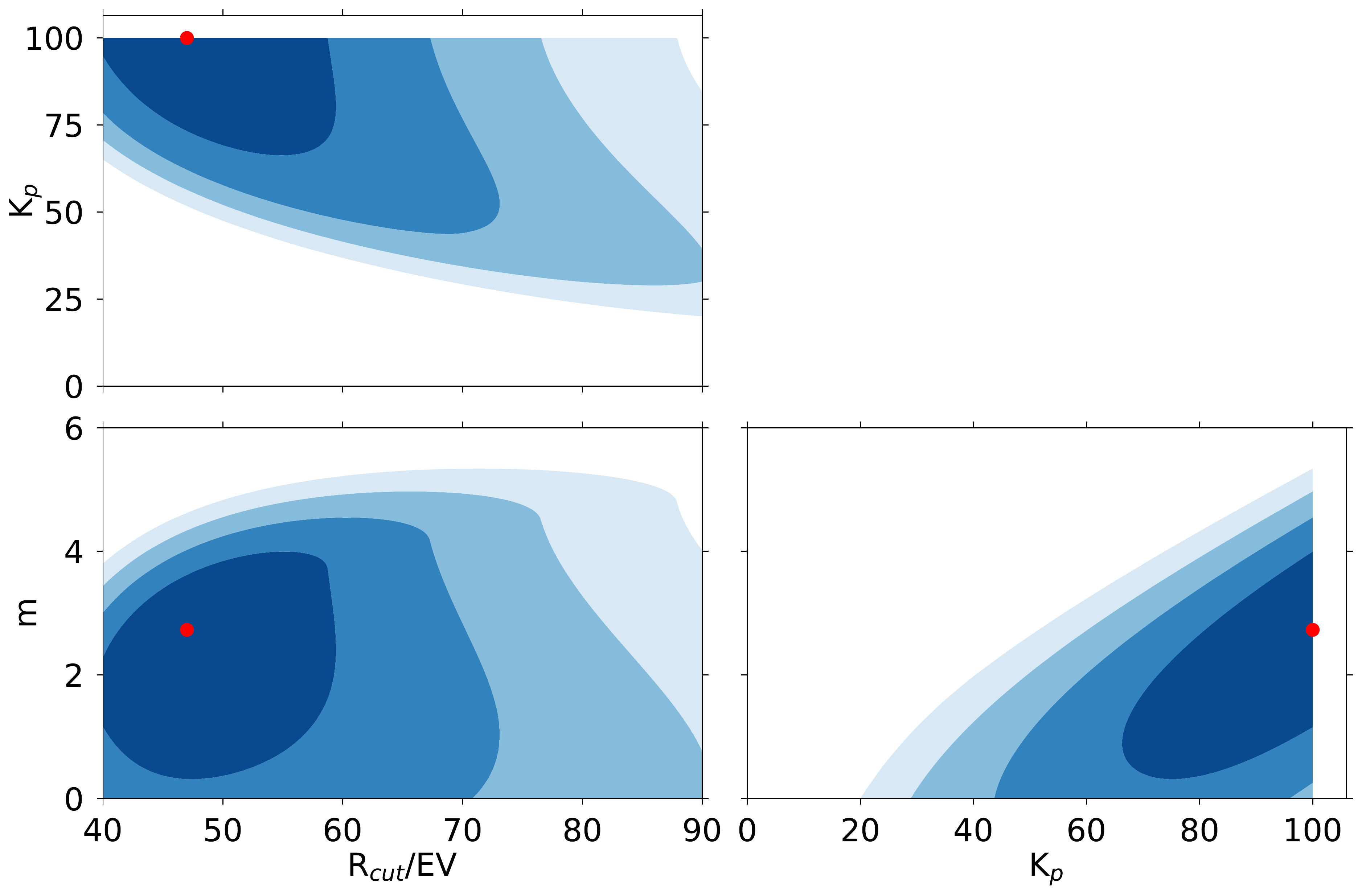}
\caption{\small{Correlation of fit parameters and best-fit values (indicated by a red dot). {\it{Top}}: $\alpha=2.2$, $z_{\text{max}}=3$. {\it{Middle}}: $\alpha=2.2$, $z_{\text{max}}=4$. {\it{Bottom}}: $\alpha=2.4$, $z_{\text{max}}=3$. The 4 shaded regions from dark to light blue are the intervals for $1\sigma$, $2\sigma$, $3\sigma$ and $4\sigma$ standard deviations.}}
\label{fig:fit_corr}
\end{figure}

\begin{figure*}
\centering
\begin{subfigure}{0.5\textwidth}
  \includegraphics[width=8.0 cm, height = 5.2 cm]{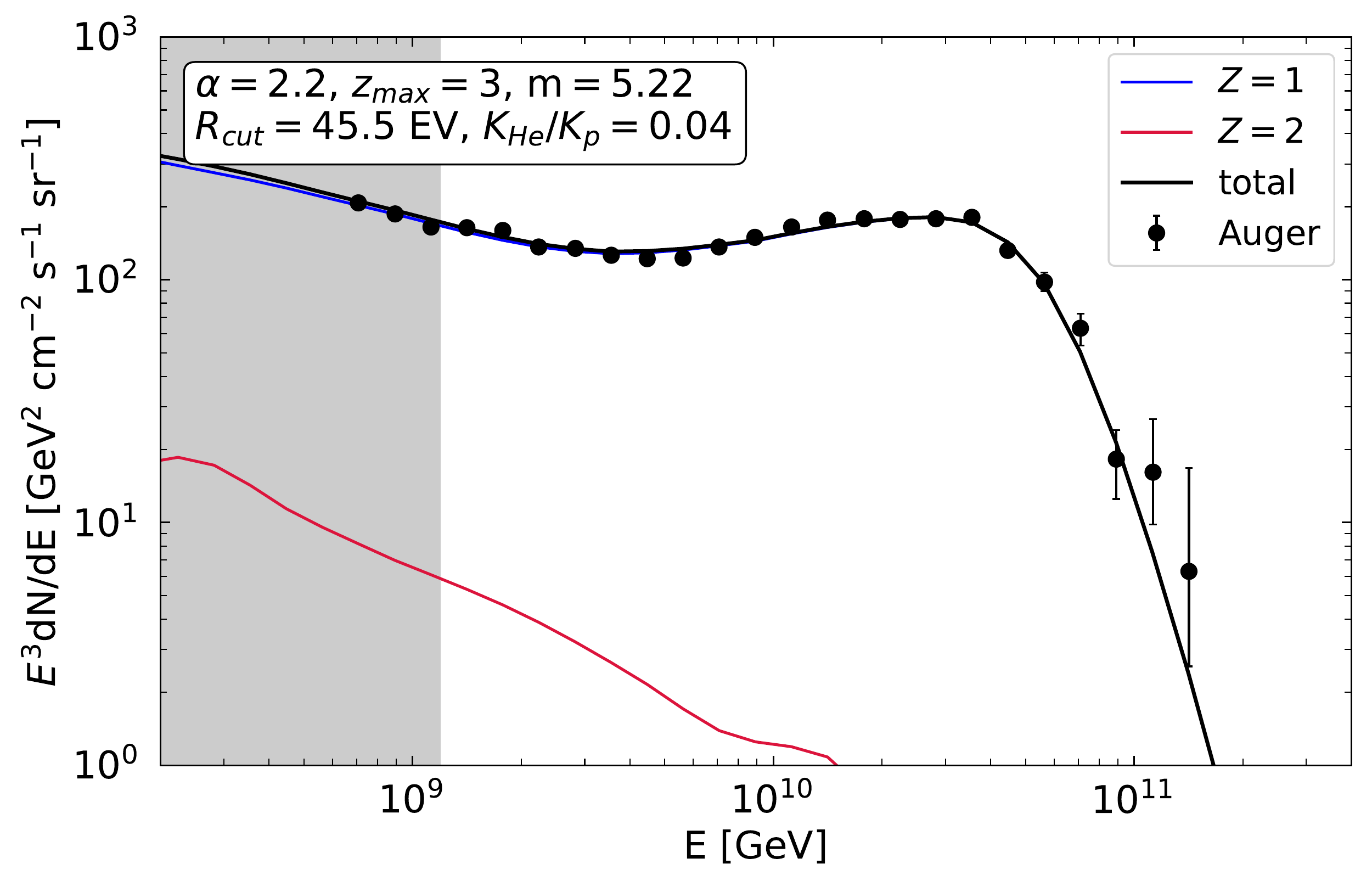}
 \end{subfigure}%
 \begin{subfigure}{0.5\textwidth}
\centering
  \includegraphics[width=8.0 cm, height = 5.2 cm]{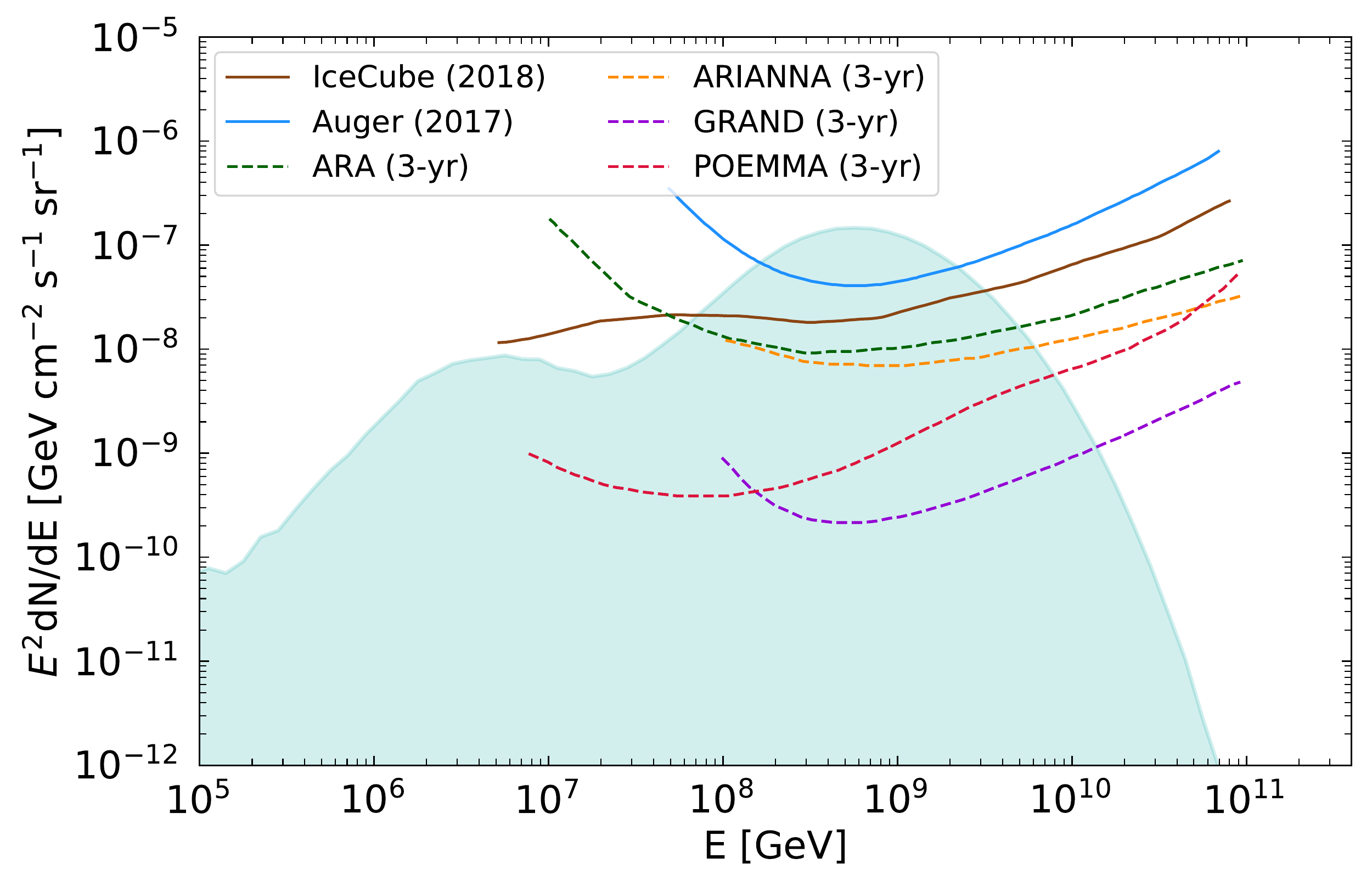}
\end{subfigure}
  \caption{\small{UHECR spectrum ({\it{left}}) and cosmogenic neutrino flux ({\it{right}}) for the best-fit case corresponding to $\alpha=2.2$ and $z_{\text{max}}=3$, found by scanning over a wide range of parameter space}}
\label{fig:a2}
\end{figure*}

We study the sensitivity of best-fit to variation of parameters $R_{\text{cut}}$, $m$, $K_{\rm p}$ and $K_{\rm He}$ for fixed values of $\alpha$ and $z_{\text{max}}$. Some general trends can be easily noted from the coarser variation of parameters given in Appendix~\ref{app:best}. For all the $\alpha$ values considered in this study, i.e., $2.2$, $2.4$, and $2.6$, the variation of $z_{\text{max}}$ within the range $2-4$ has a very little effect on the best-fit UHECR spectrum. For $\alpha=2.2$ and $2.4$, the fit improves ($\chi^2$ value decreases) with increase in source evolution index $m$. Whereas for $\alpha=2.6$, a good fit is obtained for only $m=0$ and the fit worsens for higher values of $m$. Thus, the source evolution index is found to play a major role in fitting the UHECR spectrum to Auger data (see Subsec.~\ref{subsec:uhecr}). As the injection spectral index increases, the best-fit composition approaches to pure proton for lower $m$ values, until for $\alpha=2.6$, where the best-fit composition corresponds to $100$\% proton and $m=0$. All this motivates us to explore the plausible range of parameter space and to put restrictions on them based on composition measured by PAO. Additional constraints can also be drawn to the allowed range of parameter values, whenever there is a tension with neutrino flux upper limit measured by detectors like IceCube and Auger.

To study the dependency of one parameter on the other, we vary $R_{\text{cut}}$ from 40 EV to 100 EV in steps of 0.5 EV; $m$ from 0 to 6 in steps of 0.03. At the same time, we scan over the composition space for proton and helium fraction from 0\% to 100\% with a spacing of 1\%. This gives us a total of $121\times 201 \times 101$ grid points in three-dimensional parameter space. As the variation of each parameter is computationally expensive, we restrict ourselves to discrete values of \{$\alpha$, $z_{\text{max}}$\} and discuss qualitatively the effects introduced by a variation of these. We study the relative dependency between parameters for three combinations of $\alpha$ and $z_{\text{max}}$. The best-fit values of $R_{\text{cut}}$, $m$, $K_{\rm p}$ and $\chi^2$ for those combinations are listed in Table~\ref{tab:correlation}. Fig.~\ref{fig:fit_corr} shows the variation of these parameters in a corner plot. The top, middle and bottom panel shows the variation for $\alpha=2.2$, $z_{\text{max}}=3$; $\alpha=2.2$, $z_{\text{max}}=4$; and $\alpha=2.4$, $z_{\text{max}}=3$ respectively. Here, we effectively vary three physical parameters, as the value of $K_{\rm p}$ uniquely fixes the value of $K_{\rm He}$ through the condition, $K_{\rm p} + K_{\rm He}=100$\%. For this reason, the variation with respect to helium fraction is not shown in the figures. Considering the normalization of the simulated spectra as an additional parameter, the number of degrees of freedom, $N_{\rm d}= 21 - 3 - 1 = 17$, since we fit our simulated spectrum to 21 Auger data points. The four shades from dark to light blue in the figures indicate the Bayesian confidence intervals corresponding to $1\sigma$, $2\sigma$, $3\sigma$ and $4\sigma$ standard deviations for 17 d.o.f. The best-fit values are denoted by a red dot in the plots.

\begin{table}
\caption{\label{tab:correlation}Best-fit values in parameter space [p+He] and in the energy range $E>10^{18}$ eV}
\begin{ruledtabular}
\begin{tabular}{ccccrcc}
 $\alpha$ & $z_{\text{max}}$ & $m$ & $R_{\text{cut}}$ & $K_{\rm p}$ & $K_{\rm He}$ & $\chi^2$ \\
\hline \\
$2.2$ & $3$ & $5.22$ & $45.5$ EV & $96$\% & $4$\% & $8.29$\\
$2.2$ & $4$ & $5.31$ & $45.5$ EV & $96$\% & $4$\% & $7.04$\\
$2.4$ & $3$ & $2.73$ & $47.0$ EV & $100$\% & $0$\% & $12.01$\\
\end{tabular}
\end{ruledtabular}
\end{table}

For $\alpha=2.2$, the best-fits correspond to a very strong source evolution and a composition very near to pure proton. But the constraints from neutrino fluxes alone disfavors this scenario. For such a high evolution index and $96$\% proton composition, the cosmogenic neutrino flux exceeds the flux upper limit by IceCube \citep{aartsen18} as shown in the right panel of Fig.~\ref{fig:a2} (for $z_{\text{max}}=3$). This resonates with the fact already studied in \citep{gelmini11, supanitsky16} with an older version of Auger data, that pure proton composition cannot explain the UHECR spectrum as it overproduces the neutrino and $\gamma-$ray flux. Although the fit to Auger data for UHECR spectrum turns out to be extremely good (shown in the left panel of Fig.~\ref{fig:a2}), the composition is very near to pure proton. This is in direct contradiction to the composition measurements by Auger. The standard deviation in the shower depth distribution, $\sigma(X_{\rm max})$ indicates a composition that lies between p and He up to $10^{19.5}$ eV. Hence, we reject the best-fit parameter sets obtained for $\alpha=2.2$. Instead, we restrict ourselves to $m\leqslant3$; otherwise, the He fraction gets reduced significantly at high energies, and the neutrino flux is overproduced. Comparing the cases in the top and middle panel of Fig.~\ref{fig:fit_corr}, it can be seen that only the best-fit value of source evolution index changes by a negligible amount due to a variation in $z_{\text{max}}$, within the precision adopted for this study. However, the secondary neutrino flux increases due to an increase of $z_{\text{max}}$.

Increasing $\alpha$ value decreases the transition energy between galactic and extragalactic UHECRs. For $\alpha=2.4$ the best-fit composition corresponds to pure proton. The best-fit $m$ value comes out to be 2.73. But, to avoid a pure proton composition for the same reasons as mentioned above, we restrict our parameter range for $m$ to be $\leqslant2$ for $\alpha=2.4$. In that case, the neutrino fluxes are also within the flux upper limit by IceCube as shown in Fig.~\ref{fig:2.4}. This constraint allows a significant fraction of helium to contribute to the UHECR mass composition at high energies. For $\alpha=2.6$, it is seen in Table~\ref{tab:bestfit}, that any value of $m\geqslant0$ results in a pure proton composition and the lowest $\chi^2$ occurs for $m=0$. We checked that no best-fit could be obtained within $1\sigma$ for the range of parameter values considered here. Since we confine ourselves to only positive source evolution for p+He composition, the best-fit value of $m$ for $\alpha=2.6$ cannot be limited any further, and there is no scope to add helium to the injected mass. So, $\alpha=2.6$ cases are disfavored as a plausible scenario, and the fit worsens rapidly with an increase in the value of $m$. Also, the $\chi^2$ values obtained in Table~\ref{tab:bestfit} for $\alpha = 2.6$ cases suggest that the spectral fit is poor compared to those for other $\alpha$ values. 

Note also the large contours around the best-fit positions in the parameter space in Fig.~\ref{fig:fit_corr}, which would allow the parameters that we have fixed in Table~\ref{tab:bestfit} within $1\sigma$ or $2\sigma$ confidence regions for most of the cases. For example, the best-fit case 12 shown in Fig.~\ref{fig:2.2}, bottom panel, can be directly compared with the middle panel of Fig.~\ref{fig:fit_corr}. For case 12, $m=3$, $K_p = 0.74$ and $R_{\rm cut} = 60$ EV are all within $1\sigma$ contours of the best-fit position.

\subsection{\label{subsec:photons}Constraints from cosmogenic photons\protect}

\begin{figure}
\centering
\includegraphics[width = 0.48\textwidth]{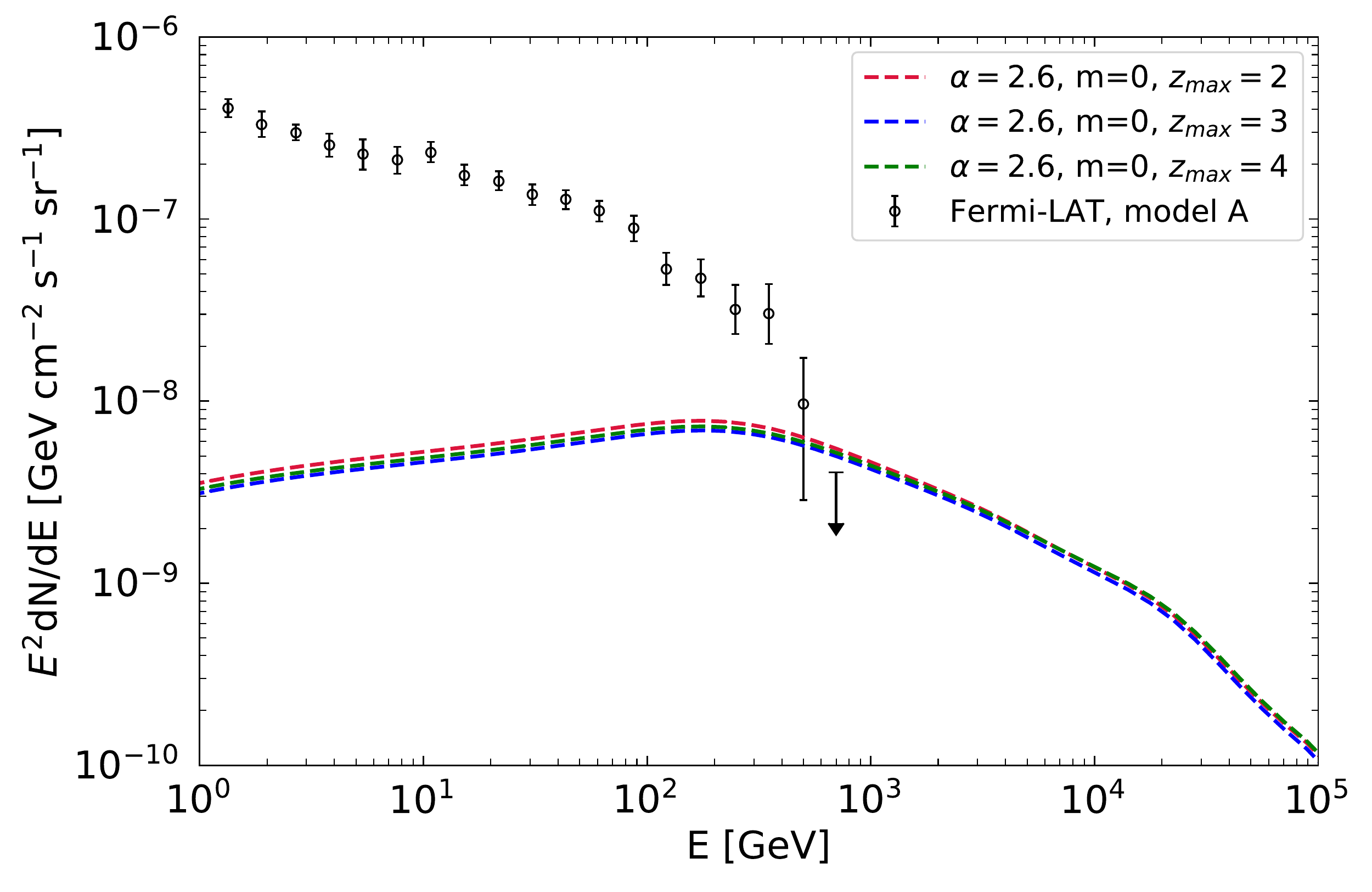}
\caption{\small{Cosmogenic photon fluxes for the $m=0$ best-fit cases, with $\alpha=2.6$ and $z_{\rm max}=2$, $3$, and $4$}. The measured diffuse gamma-ray background by Fermi-LAT is also shown.}
\label{fig:photon_2.6}
\end{figure}

The major components of cosmogenic photon flux in case of pure proton composition of UHECRs arise from the electromagnetic cascade of $\gamma-$rays produced in the decay of photopions and $\mathrm{e^+}\mathrm{e^-}$ pairs produced in Bethe-Heitler process on CMB and EBL \citep{berezinsky11, berezinsky16, murase10}. Helium or heavier elements can also produce $\gamma-$photons via photodisintegration on background photons, as it is the dominant energy loss process for them. But, the contribution to cosmogenic photon flux from photodisintegration is less compared to $p\gamma$ interactions, as discussed in \citep{stecker69, stecker99, anchordoqui07}. The Fermi-LAT measurements of the diffuse isotropic gamma-ray background (DGRB) put an upper bound to the cosmogenic photon flux produced by UHECRs. We have obtained the best-fit cases with $\alpha=2.6$ for pure proton composition and $m=0$. The spectrum for any $m$ value beyond this is strongly disfavored based on the $\chi^2$ values obtained in our work. It can be seen in Fig.~\ref{fig:photon_2.6} that the photon flux obtained for $m=0$ cases are comparable to the flux measured by Fermi-LAT at the highest energy bin corresponding to 820 GeV \citep{ackermann15}. This is in agreement with the results in \citep{berezinsky11}, which shows that only $m=0$ is compatible with the Fermi-LAT data for $\alpha=2.6$. But, this scenario is disfavored due to a pure proton composition. It is shown in \citep{batista18}, the effects on the photon flux due to change in $z_{\text{max}}$ is negligible for both $m=0$ and $3$. In this work too, we find in Fig.~\ref{fig:photon_2.6} that the photon fluxes are almost similar for $z_{\text{max}}$ values 2, 3 and 4. It has also been shown earlier that with increasing values of $\alpha$ and $m$, the cosmogenic photon flux increases \citep{berezinsky11}.

A pure proton composition with $\alpha = 2.4$, produces the maximum cosmogenic photon flux allowed by Fermi LAT measurements for $z_{\rm max} = 3$ and $m = 3$, a result found previously (see Table-1 of \citep{berezinsky11}). In our analysis, $m=2$ is the highest source evolution allowed for $\alpha=2.4$ to avoid a pure proton composition as explained in Subsec.~\ref{subsec:correlation}. The addition of helium further reduces the flux of cosmogenic photons. These two factors taken together generate a cosmogenic photon flux in our work for $\alpha=2.4$ lower than that given in \citep{berezinsky11} and Fermi-LAT measurements. Thus our best-fit parameter range $m=0-2$ and $z_{\text{max}}=2-4$ for $\alpha=2.4$ remains viable. In all $\alpha=2.2$ best-fit cases the helium fraction is very high. As a result, cosmogenic photon flux is expected to be low compared to the earlier cases. However, for $z_{\text{max}}=4$ and $m=3$, we find the peak in the neutrino flux touches the IceCube upper limit \citep{aartsen18}. Hence we do not consider this case as favorable.

\subsection{\label{subsec:heavy}Effects due to injection of heavier elements\protect}

\begin{figure*}[hbt]
\centering
\begin{subfigure}{.5\textwidth}
  \centering
  \includegraphics[width=8.0 cm, height = 5.2 cm]{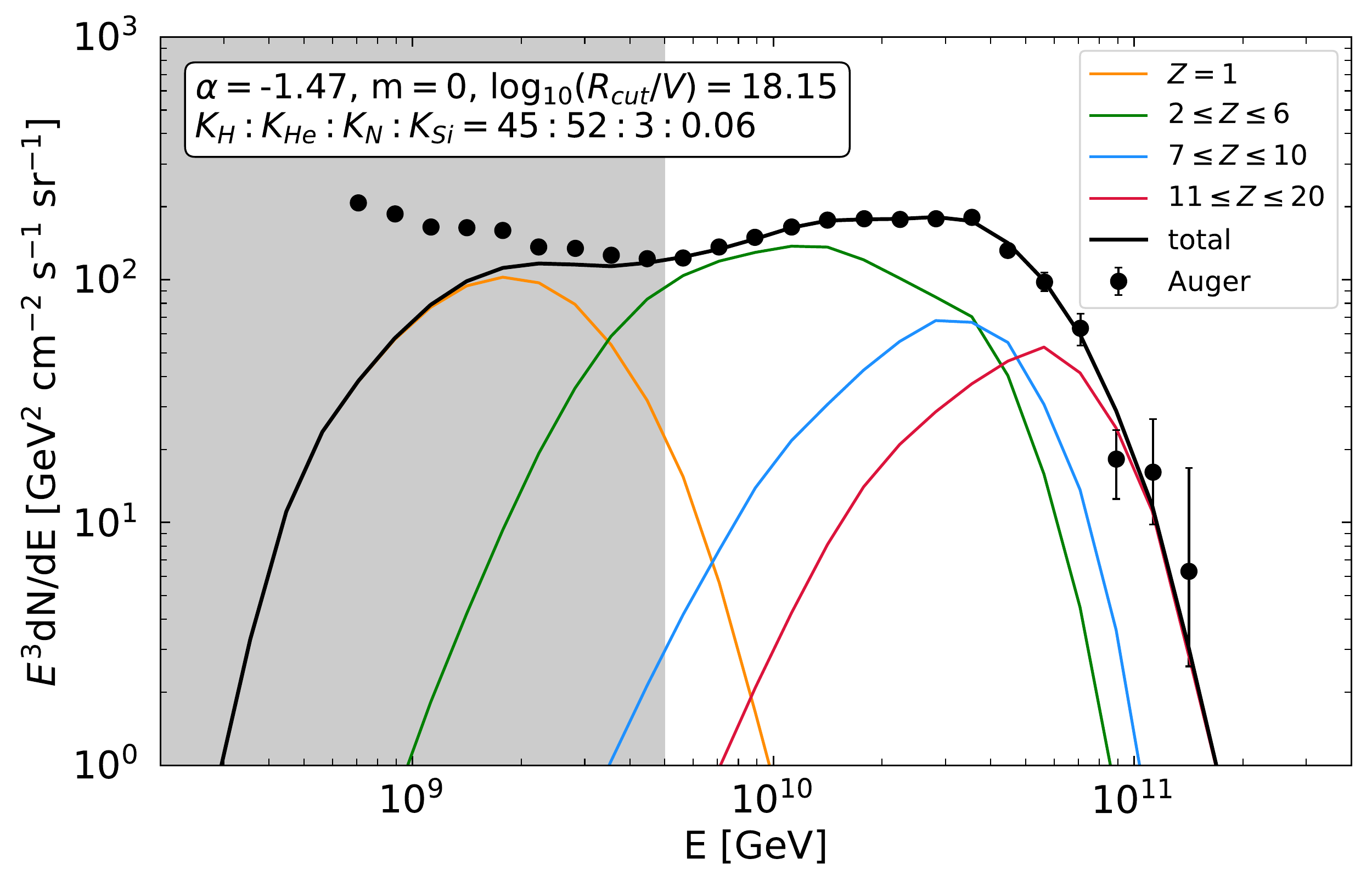}
\end{subfigure}%
\begin{subfigure}{.5\textwidth}
  \centering
  \includegraphics[width=8.0 cm, height = 5.2 cm]{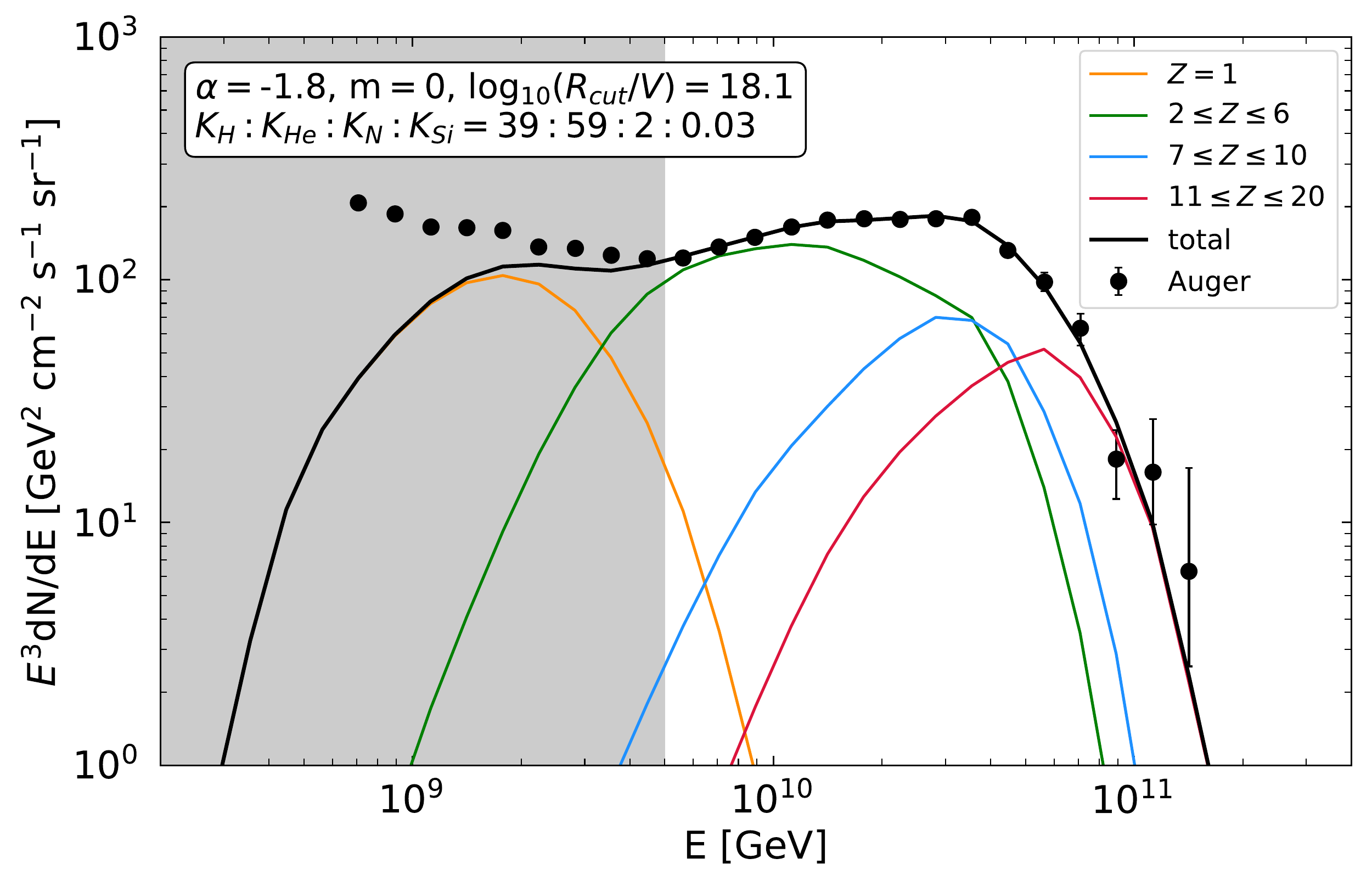}
\end{subfigure}
\begin{subfigure}{.5\textwidth}
  \centering
  \includegraphics[width=8.0 cm, height = 5.2 cm]{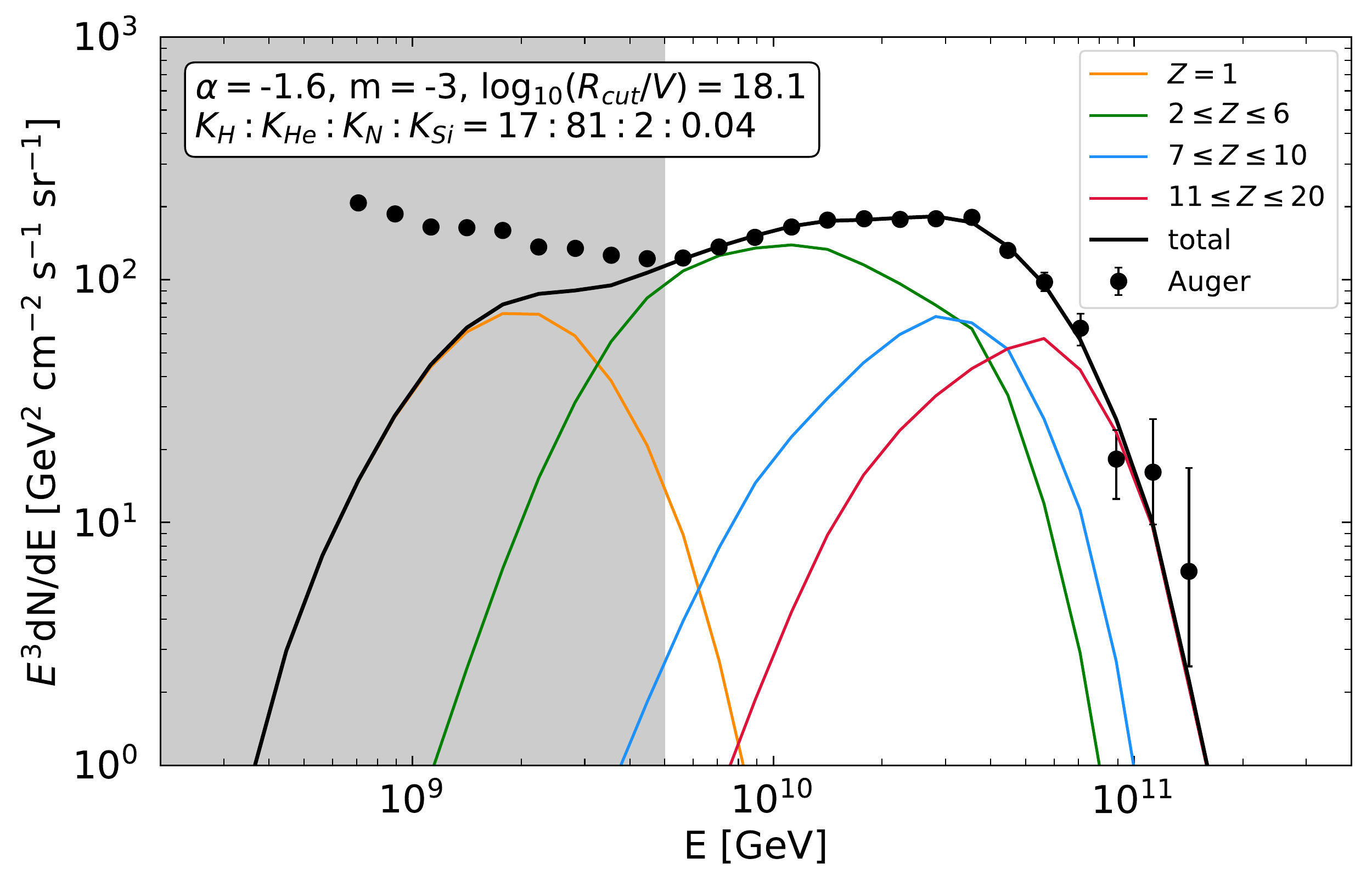}
\end{subfigure}%
\begin{subfigure}{.5\textwidth}
  \centering
  \includegraphics[width=8.0 cm, height = 5.2 cm]{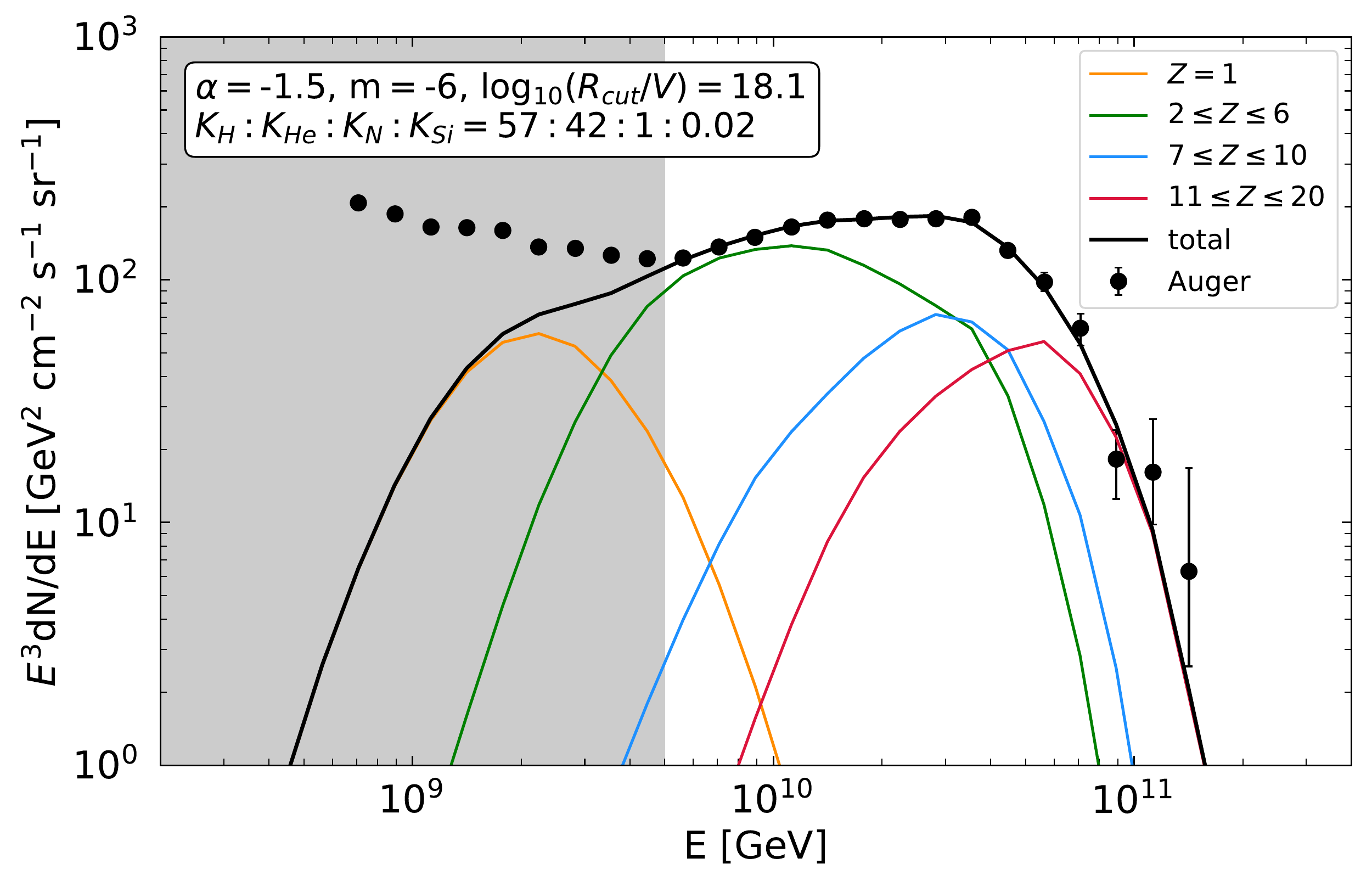}
\end{subfigure}
\caption{\small{UHECR spectra for the best-fit parameters of CTD model as found by PAO for $m=0$ (top left), and that calculated in this work for $m=$ 0, $-3$, $-6$ by extending the range of $\alpha$ used to scan the parameter space. The top right, bottom left and bottom right spectrum corresponds to $m=0$, $m=-3$ and $m=-6$ respectively as indicated in the figure labels.}}
\label{fig:auger_CTD}
\end{figure*}

We also study the effects due to the addition of heavier elements at injection on the fit of the UHECR spectrum for the CTD propagation model. We consider a mixed composition similar to Auger, consisting of stable nuclei: H, He, N, and Si at injection, which is a representative subset of injected masses. For all the propagation models studied by PAO, the contribution of iron ($^{56}$Fe) is found to be zero. In this case, a fit can be obtained for only the super-ankle region ($> 10^{18.7}$ eV), implying contribution from a different class of sources is required to explain the region below $\sim 10^{18.7}$ eV and to have heavier elements dominate at highest energies. PAO indicates that the differences among various propagation models with different physical assumptions are much larger than the statistical errors on the parameters \citep{Aab17}. PAO has found the best-fit values of the parameters for the ``CTD'' propagation model as, $\alpha=-1.47$ and $R_{\text{cut}}=10^{18.15}$ eV, respectively for $m=0$. The uncertainty in the best fit value of $\alpha$ extends down to $\alpha=-1.5$, the lowest value they have considered. They report the best-fit composition for this case as $K_{\rm H}:K_{\rm He}:K_{\rm N}:K_{\rm Si}=45:52:3:0.06$.

We extend the scan over $\alpha$ to values below $-1.5$, in the interval [$-2.5$, $0$]. $\log_{10}(R_{\text{cut}}/\rm V)$  is varied in the range [17.8, 18.3]. We consider grid spacings of 0.1 in $\alpha$ and 0.1 in $\log_{10}(R_{\text{cut}}/\rm V)$. Due to computational cost, we could not increase the precision any further. Whereas, in the analysis by PAO a grid spacing of 0.01 is considered for these parameters. The best-fit parameter set for $m=0$, obtained by PAO and in our study differs only slightly because of the dissimilarity in grid spacings and extended range of $\alpha$. However, the spectrum in both cases is found to be identical (see Fig.~\ref{fig:auger_CTD}). Again, we consider the evolution of sources in redshift is $\propto (1+z)^m$. We assume that particles are injected by sources with energies between 0.1 EeV and 1000 EeV. The injection spectrum, in this case, is the same as given in Eq.~\ref{eq:injection}. Since for heavier elements, only the particles originating from $z\lesssim 0.5$ are able to reach Earth with $E>10^{18.7}$ eV, we consider $z_{\text{max}}=1$ in the simulations. Beyond $z_{\text{max}}=1$, source evolution becomes insignificant, and a flat source evolution can be considered. Also, to cover the highest energy data points, we take $z_{\text{min}}=0$. This choice has been made by other authors as well \citep{batista18} although it is unrealistic for astrophysical source distributions. To have more realistic values of $\alpha$, i.e., to approach towards a softer injection spectrum, we find that negative source evolution is favorable for this scenario. A dense near source distribution results in higher values of $\alpha$. We find the best-fit values of composition, spectral index and $\log_{10}(R_{\text{cut}}/\rm V)$ for $m = 0, -3, -6$ and the same are listed in Table~\ref{tab:heavy}. We show the UHECR spectra in Fig.~\ref{fig:auger_CTD} for the best-fit parameters found by PAO ($m=0$) and that for the cases listed in Table~\ref{tab:heavy}.

\begin{table}
\caption{\label{tab:heavy}Best-fit values in parameter space [H+He+N+Si] and in the energy range $E>10^{18.7}$ eV}
\begin{ruledtabular}
\begin{tabular}{rccccccc}
$m$ & $\alpha$ & $\log_{10}(R_{\text{cut}}/\rm V)$ & $K_{H}$ & $K_{\rm He}$ & $K_{\rm N}$ & $K_{\rm Si}$ & $\chi^2$\\
\hline \\
$0$ & $-1.8$ & $18.1$ & $39$ & $59$ & $2$ & $0.03$ & 2.59 \\
$-3$ & $-1.6$ & $18.1$ & $17$ & $81$ & $2$ & $0.04$ & 2.57 \\
$-6$ & $-1.5$ & $18.1$ & $57$ & $41$ & $1$ & $0.02$ & 2.66\\
\end{tabular}
\end{ruledtabular}
\end{table}

\begin{figure*}[hbt]
\centering
\begin{subfigure}{.5\textwidth}
  \centering
  \includegraphics[width=8.6 cm, height = 5.2 cm]{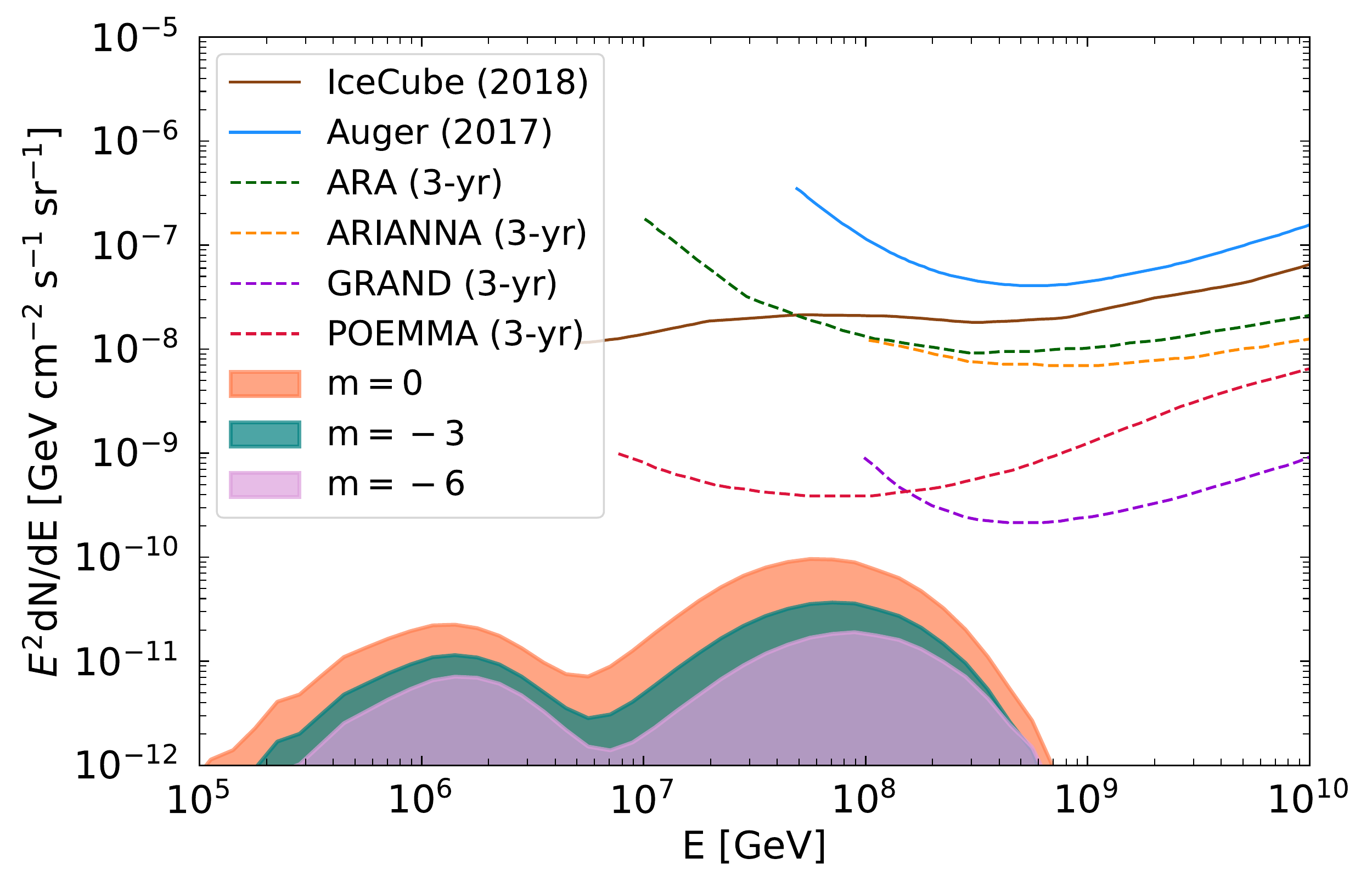}
\end{subfigure}%
\begin{subfigure}{.5\textwidth}
  \centering
  \includegraphics[width=8.4 cm, height = 5.2 cm]{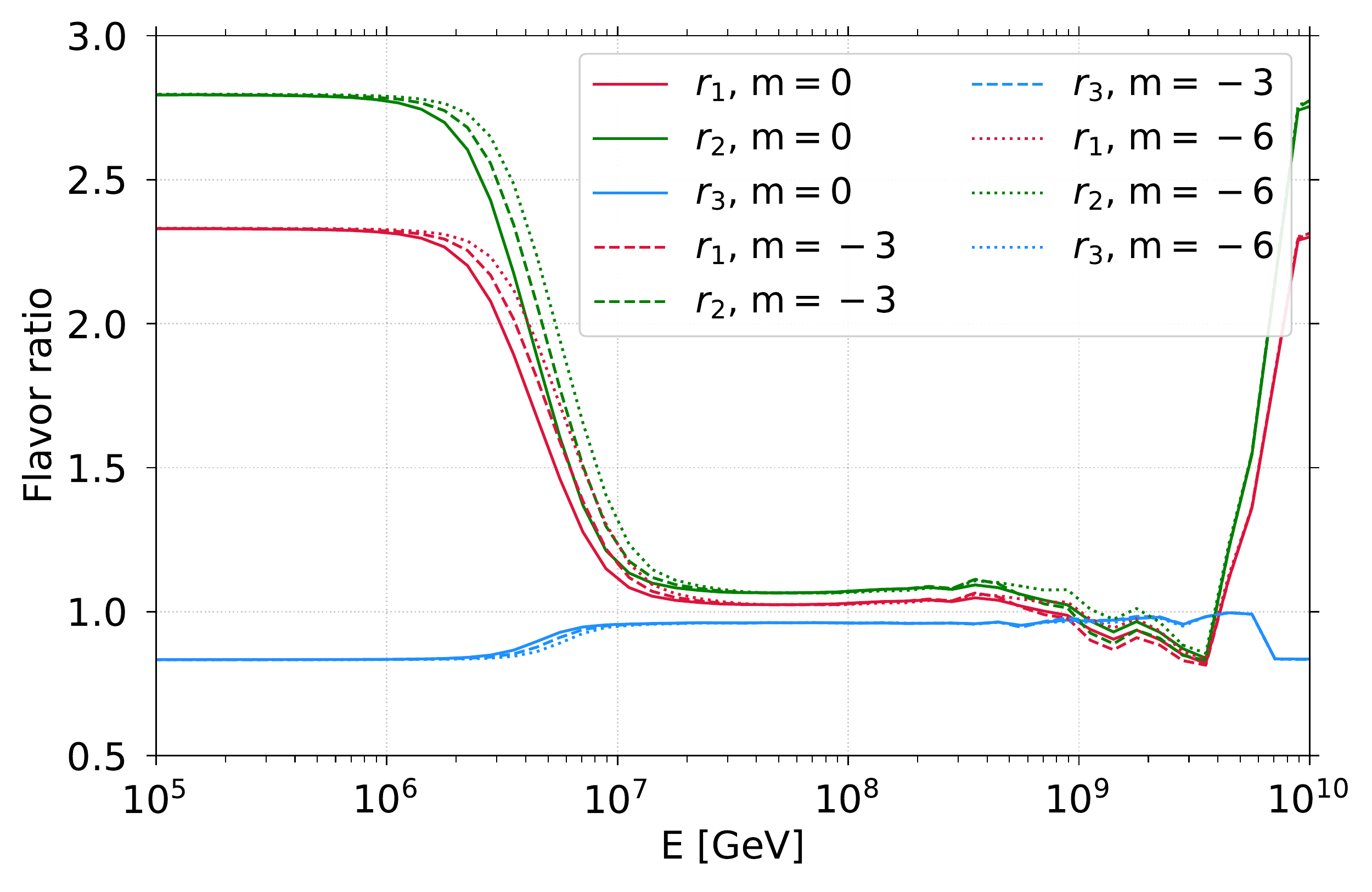}
\end{subfigure}
\caption{\small{{\it{Left}}: Cosmogenic neutrino fluxes for the best-fit parameters of CTD model, as listed in Table~\ref{tab:heavy}. {\it{Right}}: The ratio of neutrino flavor components for the evolution cases studied.}}
\label{fig:auger_neu}
\end{figure*}

The cosmogenic neutrino fluxes for the best-fit cases obtained in our study and the ratio of neutrino flavor components obtained at Earth are shown in Fig.~\ref{fig:auger_neu}. We find that the neutrino fluxes decreases with increasing negative source evolution index. This is expected because, with more number of sources at low redshift, most of the UHECRs propagate over distances shorter than the interaction length for $p\gamma$ interaction, thus reducing the secondary cosmogenic fluxes. The neutrino fluxes obtained are too small to be detected by any future neutrino detectors. The maximum flux for $m=0$ is $9.6\times10^{-11}$ GeV cm$^{-2}$ s$^{-1}$ sr$^{-1}$ at $\sim 60$ PeV. This value is far below the 3-yr integrated sensitivity of proposed detectors like GRAND and POEMMA and is not likely to be detected in the near future. The contribution to cosmogenic neutrino flux is $\lesssim 10^{-12}$ GeV cm$^{-2}$ s$^{-1}$ sr$^{-1}$ in the ultrahigh energy range ($>10^{18}$ eV). We represent the flavor ratios as $r_{e/\mu} = r_1$, $r_{e/\tau} = r_2$ and $r_{\tau/\mu} = r_3$ in the right panel of Fig.~\ref{fig:auger_neu}. We note that although there is no significant difference in the ratios for different source evolution index, there is, in general, a sharp increase in the electron neutrino flux below $10^7$ GeV and at above $3\times 10^9$ GeV. This sharp increase is primarily due to beta-decay neutrinos.

There are some difficulties involved with the injection of heavier elements. These are:
\begin{itemize}
\item To have N and Si dominate at higher energies, the injection spectral index must be too hard. This cannot be explained by the typical Fermi acceleration process.
\item With more and more negative values of $m$, the value of $\alpha$ increases to as low as $\alpha=-1.5$ for $m=-6$. Such strong negative source evolution cannot be explained by the local AGN density or any other ultraluminous sources, within the redshift range considered.
\item Any value of $z_{\text{min}}>0$ will be insufficient to accommodate the highest energy events, with the mixed composition considered here. Centaurus A is the nearest AGN known with redshift $z=0.0007$. Even with this value of $z_{\text{min}}$, the spectrum cannot cover the highest energy data points.
\item The low value of maximum rigidity increases UHECR interactions on EBL than on CMB, for medium atomic number elements (CNO) \citep{Batistaebl15}. Photodisintegration on EBL can incorporate large uncertainties in extragalactic propagation since the cross sections are poorly known.
\item It is highly improbable to generate the ankle feature using a mixed composition containing heavier elements. Thus an additional sub-ankle component of unknown nature and different physical parameters is required. A galactic contribution is also ruled out at such high energies.
\end{itemize}

In \citep{batista18} a possible explanation for the hard spectral index is given by making a distinction between the spectrum of accelerated particles and the one of escaping particles. But, the extent to which the spectrum can be hardened by such interactions of UHECRs with the ambient medium near the source is not well understood. The infrared peak of the EBL models affects the propagation of heavy nuclei strongly. A lower rigidity cutoff, in addition to a hard spectral index, is preferred in this case to avoid overproduction of secondary protons \citep{Aab17}. The lower value of $R_{\text{cut}}$ decreases the maximum energy of secondary protons and thus allows heavier elements to dominate at higher energies, requiring negative values of $\alpha$. The uncertainties in the EBL model, various air-shower models, and cross section for photodisintegration of medium nuclei on EBL translates into a considerable uncertainty in determining the mass composition \citep{Batistaebl15, heinze19, biehl18}. In \citep{taylor15}, it is shown that with an increase in the number of sources at low redshifts ($m<0$), softer injection spectra consistent with Fermi acceleration ($\alpha \simeq 2$) is obtained. They suggest low-luminosity gamma-ray BL-Lacertae objects as a potential candidate of UHECR acceleration. But the number density of bright BL Lacs peak at $z \simeq 1.2$, whereas a strong negative source evolution in the range $0\leqslant z\leqslant 1$ implies dense source distribution near to $z=0$.

\section{\label{sec:discuss}Discussions\protect}

In this work, we have addressed the prospects for an explanation of the UHECR spectrum with the  ``CTD'' propagation model over an energy range starting from $\approx 10^{18}$ eV with a single population of sources, that requires no additional sub-ankle component and is compatible with the most prevalent source redshift evolution $\propto (1+z)^m$, with $m\geqslant 0$ \citep{gelmini11}. The p+He mass composition at injection is also studied in \citep{aloisio17}, but considering an older set of Auger data \citep{pierre14} and with no cosmological evolution of the sources. A lower limit to the proton-to-helium ratio is also given in \citep{karpikov18} based on the study of shower depth distributions and hadronic interaction models. In our work, we find that positive source evolution in redshift is equally capable of explaining the Auger data. The fits obtained here allow us to constrain the source spectral index to lie between $2.2\leqslant \alpha \leqslant 2.6$. Outside this range the fit becomes poor, and no suitable parameter values conform with the Auger data down to $\approx 10^{18}$ eV. However, for $\alpha=2.6$, a pure proton composition is obtained for all the best-fit cases, which contradicts the composition measurements by Auger. Thus, $\alpha=2.6$ cases are disfavored. In \citep{heinze16}, it is shown that the pure proton dip model exceeds the neutrino flux upper limit by IceCube for various combinations of $\alpha$, $m$ and $E_{\rm max}$. A similar inference is given from studies of maximum possible cosmogenic photon fluxes in \citep{berezinsky11, berezinsky16, supanitsky16}. A study of luminosity and number density of steady sources show that UHECRs cannot be pure protons at $E>8\times10^{19}$ eV \citep{fang16a}. For $\alpha<2.6$, the addition of other nuclei to composition becomes indispensable.

The reference model ``SPG'' of the PAO fit \cite{Aab17} considers {\it{SimProp}} propagation with Puget, Stecker and Bredekamp model of photodisintegration \cite{Puget76} and Gilmore et al.\ EBL model \cite{gilmore12}. The best-fit parameters for this model indicate two minima. One minimum corresponds to a low value of $\log_{10}(R_{\text{cut}}/\rm V) = 18.5$ and a spectral index $\alpha \approx 1$. The other minimum corresponds to $\log_{10}(R_{\text{cut}}/\rm V) = 19.88$ and $\alpha = 2.04$. In the former case, heavier elements dominate at the highest energies, and a better fit to composition data is found by PAO. Whereas in the latter, the highest energy flux is dominated by light elements. Since, uncertainties introduced by poorly known quantities like photodisintegration cross section, EBL model, hadronic interaction model, etc. affect the determination of shower depth distribution $X_{\rm max}$, we explore the possibility of a fit to UHECR spectrum considering both scenarios.

We study the source parameters for H+He+N+Si composition at injection and find the best-fit cases for different $m$ values. A variation of the source evolution index $m$ reveals that $\alpha$ increases if $m$ is allowed to be negative. Such a negative source evolution is also suggested in \citep{taylor15}. The injection spectral index is found to be too hard, even for $m=-6$. Also, an additional class of extragalactic sources is required to fit the spectrum below the ankle in this scenario, owing to the inability of Galactic SNRs to accelerate particles to this energy. A possible extragalactic origin of the light nuclei component for $E<10^{18.7}$ eV is suggested in recent studies by virtue of increased photohadronic interactions close to the accelerator \citep{kachelreiss17, supanitsky18}.

We find the best-fit values of $R_{\text{cut}}$ in case of heavy element composition are low, and the observed spectrum steepens as a result of limited energy of accelerated cosmic rays at the sources and photodisintegration \cite{Aab17}. This is because of the energy per nucleon is much below the threshold for photopion production on CMB. Thus in case of heavier nuclei, the cosmogenic neutrino flux at EeV energy is found to be extremely low, beyond the sensitivity of future neutrino detectors. Whereas, the GZK cutoff requires the primary proton energy to be at least comparable to the threshold for pion-production with the CMB photons. Thus in case of light nuclei composition, the steepening of the spectrum is due to the GZK effect via increased energy loss of primaries and copious production of charged and neutral pions. The neutrino fluxes obtained in our calculations vary due to composition, injection spectrum and the maximum distance up to which the sources accelerate cosmic rays. We find that our cosmogenic neutrino flux predictions are compatible with the plausible range of models studied in \citep{kotera10}.

Currently operating neutrino detectors do not reach yet the necessary sensitivity level for detecting cosmogenic fluxes. The most stringent upper limits on the flux come from analyzing 9-years of IceCube data \citep{aartsen18}.  These limits are about an order of magnitude higher than the flux level expected for our $m=0$, pure proton dominated cases with the injection spectral index $\alpha = 2.6$. For harder injection indices, we find that the cosmogenic neutrino fluxes can reach the IceCube upper limits at $\sim 1$ EeV in some cases. Among the future detectors, the prospect for detection of cosmogenic fluxes is particularly good for POEMMA \citep{poemma1, poemma2} and GRAND \citep{grand1, grand2}.  These detectors, with a combined energy coverage of 10 PeV--100 EeV, will be able to probe most of our models. Detection of cosmogenic neutrino flux together with flavor identification will be crucial to constrain UHECR composition and their sources. 

There exist various cosmological evolution functions for source emissivity. The one considered in our study is a simple power-law redshift dependence. This can be attributed to the fact that no specific source type has been correlated so far with any UHECR event. Using 3$\times10^{4}$ cosmic rays with energies above $8\times10^{18}$ eV, an anisotropy in the arrival directions is detected by PAO at a significance level of $5.2\sigma$ \citep{PAO3}. The constraints on the amplitude of the dipolar component of anisotropy for 4 EeV $<$ E $<$ 8 EeV disfavors a Galactic origin only and provides no further insight on their origin. Using two distinct type of extragalactic gamma-ray emitters, viz. active galactic nuclei from the second catalog of hard Fermi-LAT sources (2FHL) and starburst galaxies from a sample that was examined with Fermi-LAT, a sky model of cosmic ray density is constructed in \citep{PAO4}. For energies above 39 EeV, the analysis by PAO indicates that the starburst model better explains the data of arrival directions, disfavoring the isotropy of UHECRs with $4.0\sigma$ confidence.

Low-luminosity GRBs are considered as potential candidates of UHECR sources. The nuclear composition and their survivability in the jets are not well constrained due to lack of observational data. A fit to the UHECR spectrum is obtained using an injection spectrum devoid of any power-law function in \citep{zhang18}. The Si-free models can explain the $X_{\rm max}$ distribution found by PAO but fails to fit the UHECR spectrum. Whereas, the Si-rich models explains the $X_{\rm max}$ data, as well as, the UHECR spectrum. However, the highest energy data points are not well covered in the fits. A high photon density inside the source, leading to nuclear cascade is considered in \citep{Boncioli19}. Depending on the source propagation model, this gives a good fit to the observed UHECR spectrum. The authors in Ref. \citep{Globus15} explored UHECR acceleration in GRB internal shocks and subsequent propagation to the Earth. They found very hard spectral indices for various nuclear species escaping the acceleration site. A few of their models fit Auger data above $10^{18.5}$ eV with increasingly heavy nuclei dominating at higher energies.

In a recent study, it is shown that the fit to Auger data obtained with $\dot{\varepsilon}=(1+z)^m$ is much better than that obtained using AGN source evolution \citep{batista18}. For the latter, the best-fit case overshoots the measured UHECR spectrum for $E<10^{18.7}$ eV and near $4\times10^{19}$ eV. It is thus discarded to abide by the composition suggested by PAO. The overshooting is also present for SFR and GRB source evolutions, but are less pronounced than AGN. In another study, it is suggested that TeV-PeV cosmic rays in a galactic halo are injected to the transrelativistic shear acceleration by black hole jets of active galactic nuclei and can be reaccelerated up to 100 EeV \citep{Kimura18}. This hypothesis makes FR I and FR II radio galaxies, and their blazar counterpart a promising source of UHECRs.

\section{\label{sec:conclu}Conclusions}

\begin{table*}[hbt]
\caption{\label{tab:bestfit} Best-fits to UHECR spectrum for p+He composition}
\begin{ruledtabular}
\begin{tabular}{c|c|ccccccc|c|c} 
 $\mathbf{\alpha}$ & $\mathbf{z_{\text{max}}}$ & $\mathbf{m}$ & $\mathbf{R_{\text{cut}} (EV)}$ & $\mathbf{K_{\rm p}}$ & $\mathbf{K_{\rm He}}$ & $\mathbf{K_{\rm He}/K_{\rm p}}$ & $\mathbf{\chi_{\rm spec}^2}$ & \textbf{Case} & \textbf{Neutrino flux} & Remarks\\
 & & & & & & & & & (GeV cm$^{-2}$ s$^{-1}$ sr$^{-1}$) & \\
\hline
 \multirow{12}{*}{2.2} & \multirow{4}{*}{2} & 0 & 80 & 1.7 & 98.3 & 57.82 & 38.41 & 1 & $1.385\times10^{-9}$ & Disfavored\\
 & & 1 & 80 & 6.6 & 93.4 & 14.15 & 25.69 & 2 & $2.347\times 10^{-9}$ &\\
 & & 2 & 80 & 13.2 & 86.8 & 6.58 & 17.06 & 3 & $4.366\times 10^{-9}$ &\\
 & & 3 & 60 & 42.7 & 57.3 & 1.34 & 12.58 & 4 & $8.704\times 10^{-9}$ &\\ \cline{2-9}
 & \multirow{4}{*}{3} & 0 & 80 & 1.3 & 98.7 & 75.92 & 36.34 & 5 & $1.488\times 10^{-9}$ & Disfavored\\
 & & 1 & 90 & 0.0 & 100.0 & undefined & 23.69 & 6 & $2.809\times 10^{-9}$ &\\
 & & 2 & 80 & 12.7 & 87.3 & 6.87 & 15.41 & 7 & $5.949\times 10^{-9}$ &\\ 
 & & 3 & 70 & 31.3 & 68.7 & 2.19 & 12.00 & 8 & $1.464 \times 10^{-8}$ &\\ \cline{2-9}
 & \multirow{4}{*}{4} & 0 & 80 & 1.3 & 98.7 & 75.92 & 37.15 & 9 & $1.530\times 10^{-9}$ & Disfavored\\
 & & 1 & 80 & 6.2 & 93.8 & 15.13 & 24.37 & 10 & $2.983\times 10^{-9}$ &\\
 & & 2 & 80 & 12.8 & 87.2 & 6.81 & 15.76 & 11 & $7.159\times 10^{-9}$ &\\
 & & 3 & 60 & 42.3 & 57.7 & 1.36 & 11.36 & 12 & $2.079\times10^{-8}$ &\\
 \hline
 \multirow{12}{*}{2.4} & \multirow{4}{*}{2} & 0 & 50 & 67.9 & 32.1 & 0.47 & 21.91 & 13 & $1.456\times 10^{-9}$ &\\
 & & 1 & 50 & 76.2 & 23.8 & 0.31 & 17.41 & 14 & $2.459\times 10^{-9}$ &\\
 & & 2 & 50 & 86.4 & 13.6 & 0.16 & 14.39 & 15 & $4.524\times 10^{-9}$ &\\
 & & 3 & 50 & 99.0 & 1.0 & 0.01 & 12.78 & 16 & $9.055\times 10^{-9}$ & Disfavored\\ \cline{2-9}
 & \multirow{4}{*}{3} & 0 & 50 & 68.6 & 31.4 & 0.46 & 20.89 & 17 & $1.595\times 10^{-9}$ &\\
 & & 1 & 50 & 77.0 & 23.0 & 0.3 & 16.73 & 18 & $2.947\times 10^{-9}$ &\\
 & & 2 & 50 & 87.5 & 12.5 & 0.14 & 14.05 & 19 & $6.301\times 10^{-9}$ &\\
 & & 3 & 50 & 100.0 & 0.0 & 0.0 & 12.72 & 20 & $1.541\times 10^{-8}$ & Disfavored\\ \cline{2-9}
 & \multirow{4}{*}{4} & 0 & 50 & 67.4 & 32.6 & 0.48 & 20.02 & 21 & $1.611\times 10^{-9}$ &\\
 & & 1 & 50 & 75.6 & 24.4 & 0.32 & 15.54 & 22 & $3.172\times 10^{-9}$ &\\
 & & 2 & 50 & 85.8 & 14.2 & 0.17 & 12.58 & 23 & $7.595\times 10^{-9}$ &\\
 & & 3 & 50 & 98.3 & 1.7 & 0.02 & 11.04 & 24 & $2.183\times 10^{-8}$ & Disfavored\\
 \hline
 \multirow{12}{*}{2.6} & \multirow{4}{*}{2} & 0 & 60 & 100.0 & 0.0 & 0.0 & 27.89 & 25 & $1.553\times 10^{-9}$ & Disfavored\\
 & & 1 & 60 & 100.0 & 0.0 & 0.0 & 36.20 & 26 & $2.456\times 10^{-9}$ & Disfavored\\
 & & 2 & 70 & 100.0 & 0.0 & 0.0 & 52.56 & 27 & $4.509\times 10^{-9}$ & Disfavored\\
 & & 3 & 90 & 100.0 & 0.0 & 0.0 & 79.96 & 28 & $8.980\times 10^{-9}$ & Disfavored\\ \cline{2-9}
 & \multirow{4}{*}{3} & 0 & 60 & 100.0 & 0.0 & 0.0 & 29.02 & 29 & $1.686\times 10^{-9}$ & Disfavored\\
 & & 1 & 60 & 100.0 & 0.0 & 0.0 & 38.72 & 30 & $2.920\times 10^{-9}$ & Disfavored\\
 & & 2 & 70 & 100.0 & 0.0 & 0.0 & 56.45 & 31 & $6.140\times 10^{-9}$ & Disfavored\\
 & & 3 & 90 & 100.0 & 0.0 & 0.0 & 85.32 & 32 & $1.464\times 10^{-9}$ & Disfavored\\ \cline{2-9}
 & \multirow{4}{*}{4} & 0 & 60 & 100.0 & 0.0 & 0.0 & 24.89 & 33 & $1.716\times 10^{-9}$ & Disfavored\\
 & & 1 & 60 & 100.0 & 0.0 & 0.0 & 32.60 & 34 & $3.168\times 10^{-9}$ & Disfavored\\
 & & 2 & 70 & 100.0 & 0.0 & 0.0 & 48.70 & 35 & $7.378\times 10^{-9}$ & Disfavored\\
 & & 3 & 90 & 100.0 & 0.0 & 0.0 & 75.93 & 36 & $2.062\times 10^{-9}$ & Disfavored\\
 \end{tabular}
 \end{ruledtabular}
 \end{table*}

UHECR mass composition depends on various factors such as redshift evolution of sources, maximum energy of primary particles and also on the injection spectrum which is determined by the acceleration mechanism. The much acknowledged choice of power-law injection with source spectral index at $\alpha \approx 2$  originates in the well-known Fermi mechanism. An analysis by PAO using data above $5\times 10^{18}$ eV favors a harder spectral index. PAO assumes a mixed composition at injection, with the precise element fractions being determined by specific propagation model, photodisintegration cross section and EBL spectrum. In this paper, we have studied a model called ``CTD", with CRPropa 3 propagation, TALYS 1.8 photodisintegration cross section and Dom\'{i}nguez et al. EBL model, assuming two types of astrophysical situation. In one case, sources inject only H and He nuclei and a fit is possible from $E\approx 10^{18}$ eV. In another case, a mixed composition of H, He, N, Si is considered at injection and a fit is possible for $E>10^{18.7}$ eV. We constrain the range of injection spectral index and the cutoff rigidity feasible in the light nuclei injection model. We have also calculated the cosmogenic neutrino fluxes from all production channels for both scenarios.

The allowed range of parameter values yields neutrino spectra consistent with the flux upper limits imposed by present detectors. This suggests that the abundance fraction of H and He considered in the best-fit cases are plausible. The ratio of fluxes of different flavors obtained on Earth after neutrino oscillation is consistent with our expectations. The $\chi^2$ value obtained for H+He model in the fitting procedure of UHECR spectrum favors $\alpha = 2.2, 2.4$ cases over the pure proton case of $\alpha = 2.6$. The source redshift evolution is found to play a significant role in determining the flux. In particular, the $\chi^2$ analysis disfavors the $m = 0$ case for $\alpha = 2.2$. While positive values of source evolution index are preferable for light nuclei composition, negative values are necessary to obtain realistic values of injection spectral index for heavier composition. 

With an increase in maximum source redshift, there is an increase in neutrino flux, due to increased propagation length of primary particles. Future neutrino telescopes with higher sensitivities at $> 1$ PeV energies will be able to probe a range of flux models we predict. A measurement will be able to constrain the maximum redshift of the UHECR source distributions. Furthermore, neutrino flavor identification will shed light on the abundance fraction of nuclei in the UHECR spectrum at injection, as shown in neutrino flavor ratios for our flux models. While we show that H+He model, as well as the H+He+N+Si composition model,  both are capable of fitting UHECR data starting from different energy values, future cosmogenic neutrino data will provide a robust test for these scenarios.

\appendix

\section{\label{app:best}UHECR parameter sets}

For each possible combination of \{$\alpha$, $z_{\text{max}}$\}, we vary $m$ and calculate the best-fit value of $R_{\rm cut}$ and composition. We list them in Table~\ref{tab:bestfit}. There are 36 cases: 12 for each value of $\alpha$. They are further subgrouped according to the maximum source redshift. For each $\alpha$, we select two cases having the lowest and highest cosmogenic neutrino flux at the higher energy peak for display. The parameter set for lowest $\chi^2$ coincides with that for maximum neutrino flux. These are cases 2 and 12 respectively for $\alpha=2.2$; cases 13 and 23 for $\alpha=2.4$; cases 25 and 33 for $\alpha=2.6$. These are shown accordingly from top to bottom in Figs.~\ref{fig:2.2}, \ref{fig:2.4} and \ref{fig:2.6} with the UHECR spectrum on the left and cosmogenic neutrino flux on the right panels. Parameter sets are labeled as ``Disfavored" following the restrictions -- only $\chi^2<27.95$ cases are accepted in this study (described in Subsec.~\ref{subsec:uhecr}), and compositions very near or equal to pure proton are in disagreement with recent measurements by PAO. Some models, such as case 12 and 24 are also disfavored by current neutrino flux upper limit from IceCube data.

\nocite{*}

\bibliography{apssamp}

\end{document}